\documentclass[fleqn,usenatbib]{mnras}

\usepackage[T1]{fontenc}

\DeclareRobustCommand{\VAN}[3]{#2}
\let\VANthebibliography\thebibliography
\def\thebibliography{\DeclareRobustCommand{\VAN}[3]{##3}\VANthebibliography}

\usepackage{graphicx}	
\usepackage{amsmath}	
\usepackage{amssymb}	
\usepackage{mathtools} 

\newcommand{\mathsout}[1]
{\bgroup\mathchoice
	{\sbox0{$\displaystyle{#1}$}%
		\usebox0\hspace{-6.1pt}%
		\rule[0.52\ht0-0.5\dp0-.5pt]{5pt}{0.6pt}}%
	{\sbox0{$\textstyle{#1}$}%
		\usebox0\hspace{-\wd0}%
		\rule[0.5\ht0-0.5\dp0-.5pt]{\wd0}{0.6pt}}%
	{\sbox0{$\scriptstyle{#1}$}%
		\usebox0\hspace{-\wd0}%
		\rule[0.5\ht0-0.5\dp0-.5pt]{\wd0}{0.6pt}}%
	{\sbox0{$\scriptscriptstyle{#1}$}%
		\usebox0\hspace{-\wd0}%
		\rule[0.5\ht0-0.5\dp0-.5pt]{\wd0}{0.6pt}}%
	\egroup}

\usepackage[normalem]{ulem} 

\usepackage{newtxtext,newtxmath}

\title[Gravitational waves from magnetars]{Continuous gravitational waves from trapped magnetar ejecta and the connection to glitches and antiglitches}

\author[G. Yim et al.]{
Garvin Yim$^{1}$\thanks{E-mail: g.yim@pku.edu.cn},
Yong Gao$^{2, 1}$,
Yacheng Kang$^{2, 1}$,
Lijing Shao$^{1, 3}$\thanks{E-mail: lshao@pku.edu.cn} and 
Renxin Xu$^{2, 1}$\thanks{E-mail: r.x.xu@pku.edu.cn}
\\
$^{1}$Kavli Institute for Astronomy and Astrophysics, Peking University, Beijing 100871, China \\
$^{2}$Department of Astronomy, School of Physics, Peking University, Beijing 100871, China \\
$^{3}$National Astronomical Observatories, Chinese Academy of Sciences, Beijing 100012, China
}

\date{Accepted XXX. Received YYY; in original form ZZZ}

\pubyear{2023}

\begin{document}
\label{firstpage}
\pagerange{\pageref{firstpage}--\pageref{lastpage}}
\maketitle

\begin{abstract}
Gravitational waves from isolated sources have eluded detection so far. The upper limit of long-lasting continuous gravitational wave emission can now probe physically-motivated models with the most optimistic being strongly constrained. Naturally, one might want to relax the assumption of the gravitational wave being quasi-infinite in duration, leading to the idea of transient continuous gravitational waves. In this paper, we outline how to get transient continuous waves from magnetars (or strongly-magnetised neutron stars) that exhibit glitches and/or antiglitches and apply the model to magnetar SGR J1935+2154. The toy model hypothesizes that at a glitch or antiglitch, mass is ejected from the magnetar but becomes trapped on its outward journey through the magnetosphere. Depending on the height of the trapped ejecta and the magnetic inclination angle, we are able to reproduce both glitches and antiglitches from simple angular momentum arguments. The trapped ejecta causes the magnetar to precess leading to gravitational wave emission at once and twice the magnetar's spin frequency, for a duration equal to however long the ejecta is trapped for. We find that the gravitational waves are more detectable when the magnetar is: closer, rotating faster, or has larger glitches/antiglitches. The detectability also improves when the ejecta height and magnetic inclination angle have values near their critical values, though this requires more mass to be ejected to remain consistent with the observed glitch/antiglitch. We find it unlikely that gravitational waves will be detected from SGR J1935+2154 when using the trapped ejecta model.

\end{abstract}




\section{Introduction}
\label{section_introduction}

With the ability to now detect gravitational waves (GWs), a new era has begun that will be filled with exciting and unexpected discoveries. So far, confident detections of GWs have come solely from compact binary coalescences - the inspiral of two compact objects such as black holes and/or neutron stars (NSs) \citep[e.g.][]{LVK2021GWTC3}. However, recent results from pulsar timing arrays have hinted at evidence for a stochastic GW background \citep{nanograv2023, epta2023, ppta2023, cpta2023}. Now that the O4 Observation Run is underway, there is renewed hope to detect some other form of GW, specifically, a GW signal that is continuous or burst-like in nature.

Each of these forms of GWs are thought to arise from different astrophysical sources \citep[e.g.][]{andersson2019GWbook} and in this paper, we will focus on continuous GWs (CWs). Traditionally, these refer to GWs that last for a quasi-infinite duration and certainly much longer than any given observation run. Possible production mechanisms include from NS mountains (i.e. a non-axisymmetric deformation of a rotating NS), precessing NSs, long-lived $r$-mode oscillations of a rotating fluid NS, accreting binary systems and even from exotic systems such as boson clouds around spinning black holes. For recent reviews of CW models and searches, see \citet{LVK2022narrowbandsearch, piccinni2022, riles2023, wette2023}.

In recent years, there has also been a keen interest on \textit{transient} CWs, that is, CWs that have a finite duration, on the order of minutes to months \citep{prixGiampanisMessenger2011}. Often, these types of transient CWs are triggered by some electromagnetically observed phenomenon. For instance, pulsar glitches, which are events where a NS suddenly increases its spin, are thought to produce transient mountains \citep{yimJones2020}, Ekman flow \citep{vanEysdenMelatos2008, bennettvanEysdenMelatos2010, singh2017} or could seed some other non-axisymmetric deformation through the excess energy released from the sudden transfer of angular momentum from the superfluid interior to the non-superfluid parts of the NS \citep{prixGiampanisMessenger2011}. Transient CWs are also expected from magnetar giant flares, short energetic bursts of hard X-ray and soft $\gamma$-ray radiation, which are expected to excite long-lived polar Alfv\'en waves \citep{kashiyamaIoka2011}. Finally, as observed for the binary NS merger GW170817, short $\gamma$-ray bursts and kilonovae arise from the coalescence of two NSs \citep{LVC2017GW170817, LVC2017GW170817multimessenger}. In such a scenario, the merger remnant could be a supramassive (uniformly rotating) or hypermassive (differentially rotating) NS, with a mass exceeding the maximum allowed for a non-rotating NS (so around 2-3~$\text{M}_\odot$). These NSs could conceivably last long enough to generate considerable transient CW radiation before collapsing into a black hole \citep{raviLasky2014, baiottiRezzolla2017}. 

In this paper, we put forward another transient CW model but here the associated observations are magnetar glitches and antiglitches, an antiglitch being a sudden spin-down of the magnetar and is a relatively rare event when compared to glitches. A magnetar is thought to be similar to a NS but with a much stronger magnetic field and is thought to be primarily powered by the evolution and decay of its magnetic field (see \citet{kaspiBeloborodov2017} for a review). One of the key features of our model is the ability to explain magnetar glitches and antiglitches simultaneously in a natural, straightforward and testable way. It is easy to imagine that the aforementioned model extends to NS glitches and antiglitches too, however, we focus on magnetars as there are recent astrophysical motivations to do so. 

Before we describe the model in Section \ref{subsection_the_trapped_ejecta_model}, we will summarise the observations that inspired this work. The magnetar of interest is SGR J1935+2154 \citep{israeletal2016} which, using X-ray timing, was found to have a spin frequency of $\nu \approx 0.308~\text{Hz}$ and a spin-down rate of $\dot{\nu} \approx - 3.5 \times 10^{-12}~\text{Hz~s}^{-1}$, though the spin-down rate can vary as much as 20\% on time-scales as short as two months \citep{younesetal2023}. It is unusual in the fact that it is, so far, one of only two magnetars to have conclusively undergone an antiglitch, with the other magnetar being 1E~2259+586 \citep{archibaldetal2013, younesetal2020}. It was also the first galactic magnetar to have been associated with a fast radio burst (FRB) \citep{CHIME2020, bocheneketal2020}.

It was recently reported by \cite{younesetal2023} that SGR J1935+2154 experienced an antiglitch on 2020 October 5. Then, three days later, FRBs were emitted from the magnetar and then less than one day after that, there was a period of pulsed radio emission observed by FAST \citep{zhuetal2020, zhuetal2023}. In total, there were 3 FRBs detected, each lasting a few milliseconds, and all three were emitted within a single rotation of the magnetar \citep{good2020}. \cite{younesetal2023} argued that given the rarity of antiglitches and FRBs, their observed synchronisation suggests an association which could help provide clues for what might trigger an FRB or pulsed radio emission. Ultimately, they proposed that the antiglitch was due to a strong short-duration particle wind which, naturally, carries angular momentum away \citep[see also][]{tong2023}. As for the generation of the FRBs, a strong wind is thought to ``comb out'' the magnetar's magnetic field lines into an almost radial configuration, resulting in an altered magnetosphere that could temporarily favour the production of FRBs and pulsed radio emission. However, it should be noted that the pulsed radio emission can only occur once the wind stops, as the opaque conditions prevent large electric potential gaps from forming which is thought to be crucial for pair production, curvature radiation and ultimately pulsed radio emission \citep{goldreichJulian1969, sturrock1971, rudermanSutherland1975}.

Two weeks after \cite{younesetal2023} first reported an \textit{antiglitch} followed by FRBs, \cite{geetal2022} announced a complementary discovery - a \textit{glitch} followed by FRBs from the same magnetar, SGR J1935+2154. Like the antiglitch case, the FRBs came three days after the glitch, though it was not reported whether there was a period of pulsed radio emission afterwards. What is clear though is that individual magnetars can exhibit both glitches and antiglitches and in both cases, they can trigger an FRB. 

We will develop a toy model, based on the idea that perhaps the particle wind is not as strong as \cite{younesetal2023} have suggested, and instead, the ejecta becomes trapped at a certain height above the (dipolar) magnetic poles, imaginably due to higher order magnetic multipoles. The rigid rotation of the trapped ejecta with the magnetar therefore results in a time-varying non-axisymmetric mass distribution which is exactly what is required for GW emission, so GWs will be given off for as long as the ejecta remains trapped. 

In the following section, we will describe the details of the model and how glitches and antiglitches emerge. Then, in Section~\ref{section_gravitational_wave_radiation}, we will look at the GWs emitted from such a system, including the detectability with future GW detectors. 
Finally, in Section~\ref{section_summary_and_discussion}, we will discuss our results and summarise our conclusions.

\section{A toy model for glitches and antiglitches}
\label{section_a_toy_model_for_glitches_and_antiglitches}

\subsection{Existing models}

The main goal of our model is to be able to explain (magnetar) glitches and antiglitches within a unified model. At least for NSs, there exists well established glitch models based on starquakes \citep{ruderman1969, baymPines1971} or superfluid vortex unpinning \citep{andersonItoh1975}. However, without a companion \citep{duccietal2015, howittMelatos2022} or some contrived mechanism, it is difficult to generate antiglitches with these standard glitch models. 

Consequently, there have been attempts to explain (magnetar) antiglitches as due to a period of enhanced particle wind \citep{tongetal2013, tong2014, younesetal2023} or a sudden decrease in the internal toroidal magnetic field strength \citep{mastranoSuvorovMelatos2015} but then the issue flips and it becomes difficult to explain glitches in such models. The particle wind models also predict simultaneous electromagnetic radiation but at least for the antiglitch in SGR J1935+2154, there was no immediate change to any electromagnetic observations \citep{younesetal2023}. 

More recently, \cite{wuZhaoWang2023} used the idea of asteroid collisions to explain magnetar glitches and antiglitches, with incoming asteroids imparting either positive or negative angular momentum to the magnetar depending on whether the impact is with or against the magnetar's rotation. Their model is neat as they are able to also explain any subsequent FRBs, however, their model requires angular momentum to be transferred at the magnetospheric radius instead of at the moment of impact and, at least to us, it is not clear why this is the case.\footnote{For instance, if a tidally-disrupted asteroid forms an accretion disk at the magnetospheric radius, it will only spin-down the magnetar as \cite{wuZhaoWang2023} themselves found that the magnetospheric radius is larger than the co-rotation radius \citep{ghoshLamb1979, andersson2019GWbook}. There is also the issue of how long it takes for the glitch to occur, the so-called ``glitch rise time''. Unless the magnetic coupling is extremely strong between the disk and the magnetar, then in the standard accretion model, it will take several years to transfer the disk's angular momentum to the magnetar \citep{ghoshLamb1979}.}

\cite{yimJones2023} also proposed a model for glitches and antiglitches where the excitation and decay of oscillation modes, propagating with or against rotation, imparts a sudden negative or positive change to the magnetar's angular momentum. Their model is more suited for small glitches, however, this means it can be immediately tested for large glitches and antiglitches with current GW detectors. A coincident glitch or antiglitch in O4 or later would allow us to put constraints on how large the oscillations are and provide limits on how much the proposed mechanism contributes.

\subsection{The trapped ejecta model}
\label{subsection_the_trapped_ejecta_model}

\begin{figure*}
	\includegraphics[width=\linewidth]{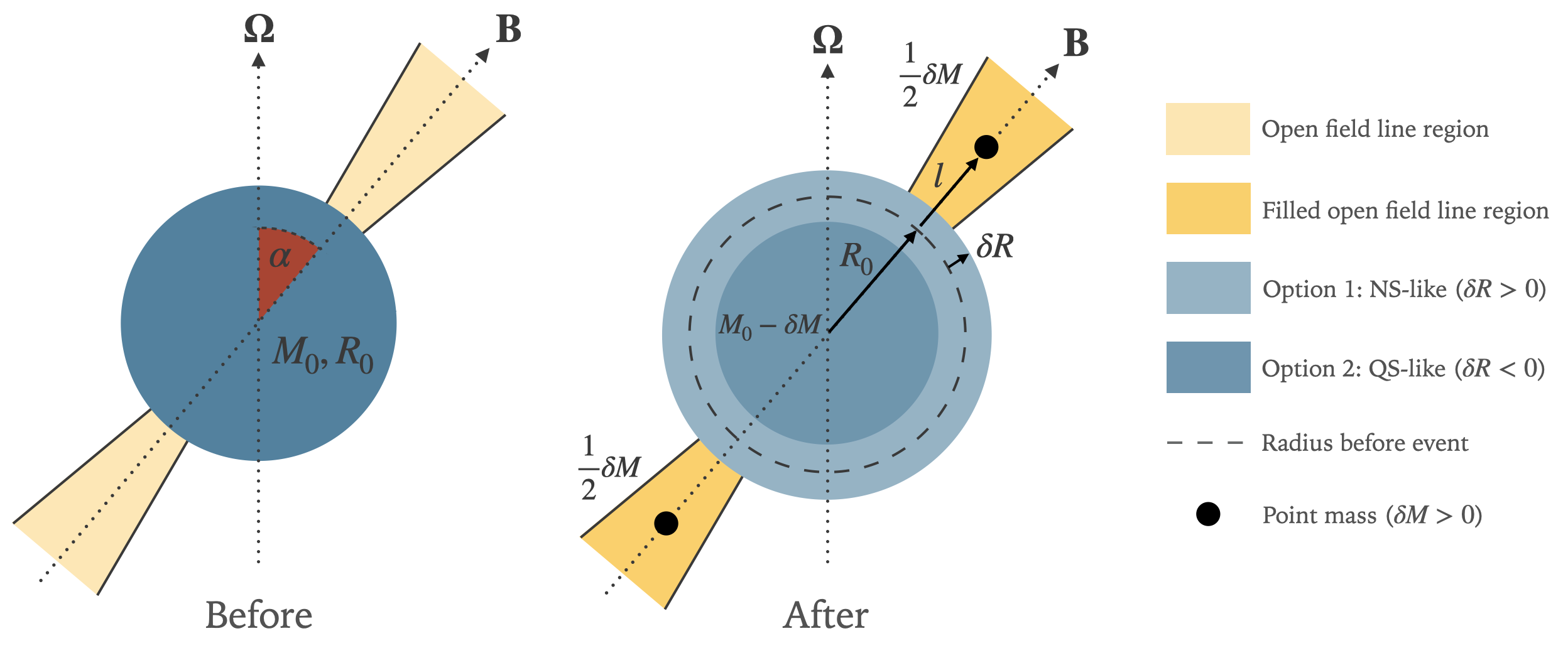}
	\caption{
		\label{fig:model_diagram}
		A diagram showing the mechanics behind the trapped ejecta model. The left diagram shows the configuration before the glitch/antiglitch event with the magnetic axis inclined by $\alpha$ from the rotational axis. The initial mass and radius of the magnetar are $M_0$ and $R_0$, respectively. Then, at the glitch/antiglitch event, the magnetar loses mass $\delta M > 0$ which gets split equally between the two magnetic poles and gets trapped at height $l$ above the pre-glitch radius. As a result of mass loss, the radius of the magnetar changes depending on whether the magnetar has a NS-like EOS ($\delta R > 0$) or a QS-like EOS ($\delta R < 0$). The decrease in moment of inertia from the magnetar and the increase in moment of inertia due to the trapped ejecta are in contest with each other and whichever dominates determines if a glitch or antiglitch occurs.
	}
\end{figure*}

In this work, we propose a novel model to simultaneously explain glitches and antiglitches which is testable with GW observations. In the remaining sections, we will not distinguish too distinctly between magnetars and NSs, but when the time comes, we will utilise arguments that are usually associated with magnetars. 

The trapped ejecta model goes as follows. Imagine a NS with mass $M_0$ and radius $R_0$ rotating with angular frequency $\Omega = 2\pi \nu$, where $\nu$ is the spin frequency of the NS. This NS is endowed with a (dipolar) magnetic field which is inclined by angle $\alpha$ from the rotation axis. This pre-glitch set-up is shown visually on the left hand side of Fig.~\ref{fig:model_diagram}.

Then, by construction, we say that the NS ejects matter out along the open field lines located above the magnetic poles and becomes trapped there. The details of the trapping are not examined in this paper but one could imagine that higher order magnetic multipoles, expected for magnetars \citep[e.g.][]{igoshevPopovHollerbach2021}, could prevent the ejecta from flowing out along open field lines. This is analogous to solar prominences, where plasma from the Sun's surface is ejected but trapped by the complex magnetic field near the surface. 

This process changes the magnetar's mass from $M_0$ to $M_0 - \delta M$, where $\delta M > 0$. By conserving mass, it means that $\frac{1}{2}\delta M$ of the ejected mass is held above each magnetic pole. We have made two assumptions here: the ejected mass is split equally between the magnetic poles and that the ejecta is trapped above the (dipolar) magnetic poles. In reality, there may be some asymmetry in the splitting of the ejected mass and the trapped ejecta might not even be antipodal from one another, or if they are, there is no guarantee that the axis connecting the antipodal ejecta aligns with the dipolar magnetic axis. Either way, we make these simplifying assumptions for now and they can be relaxed at a later date. Also, we will assume that the ejected mass is concentrated at a single point, at height $l$ above the pre-glitch radius $R_0$. It is possible to relax this assumption, but at least here, we focus more on the conceptual implications of the model, so we will not concern ourselves about the exact details of how the ejecta is trapped and what shape it takes. One can imagine these details will not affect our conclusions here too much. We also assume the ejecta rotates rigidly with the magnetar which is expected if the magnetosphere is strongly coupled to the magnetar \citep[e.g.][]{ponsVigano2019}. Regardless of the strength of the coupling, if the ejecta is held within the co-rotation radius $R_\mathrm{co}$, the radius at which the Keplerian angular frequency of the trapped matter matches the magnetar's angular frequency, then it is always possible to achieve co-rotation if there is a frictional spin-down torque acting on the ejecta. The co-rotation radius is given by
\begin{align}
R_\mathrm{co} &= \left(\frac{G M_0}{4 \pi^2 \nu^2} \right)^{\frac{1}{3}}~, \\
\label{corotation_radius}
\frac{R_\mathrm{co}}{R_0} &\approx 170 \left(\frac{M_0}{1.4~\text{M}_\odot}\right)^{\frac{1}{3}} \left(\frac{\nu}{1~\text{Hz}}\right)^{-\frac{2}{3}} \left(\frac{R_0}{10~\text{km}}\right)^{-1}~,
\end{align}
and for SGR J1935+2154 with $\nu \approx 0.308~\text{Hz}$, we find $R_\mathrm{co} \approx 370 R_0$.

Now that the magnetar has lost mass, its radius and moment of inertia must change. How the radius changes depends on whether the magnetar has a NS-like EOS or a quark star (QS)-like EOS \citep{ozelFreire2016}. Explicitly, for typical NS EOSs, the radius increases when a NS-like object loses mass, i.e. $dR/dM < 0$ (the inverse of the gradient on a mass-radius diagram). For a QS EOS, the radius decreases when a QS-like object loses mass, i.e.~$dR/dM > 0$. For our convention of having $\delta M > 0$ representing mass loss, we get $\delta R > 0$ for NSs and $\delta R < 0$ for QSs. 

\begin{figure}
	\includegraphics[width=\linewidth]{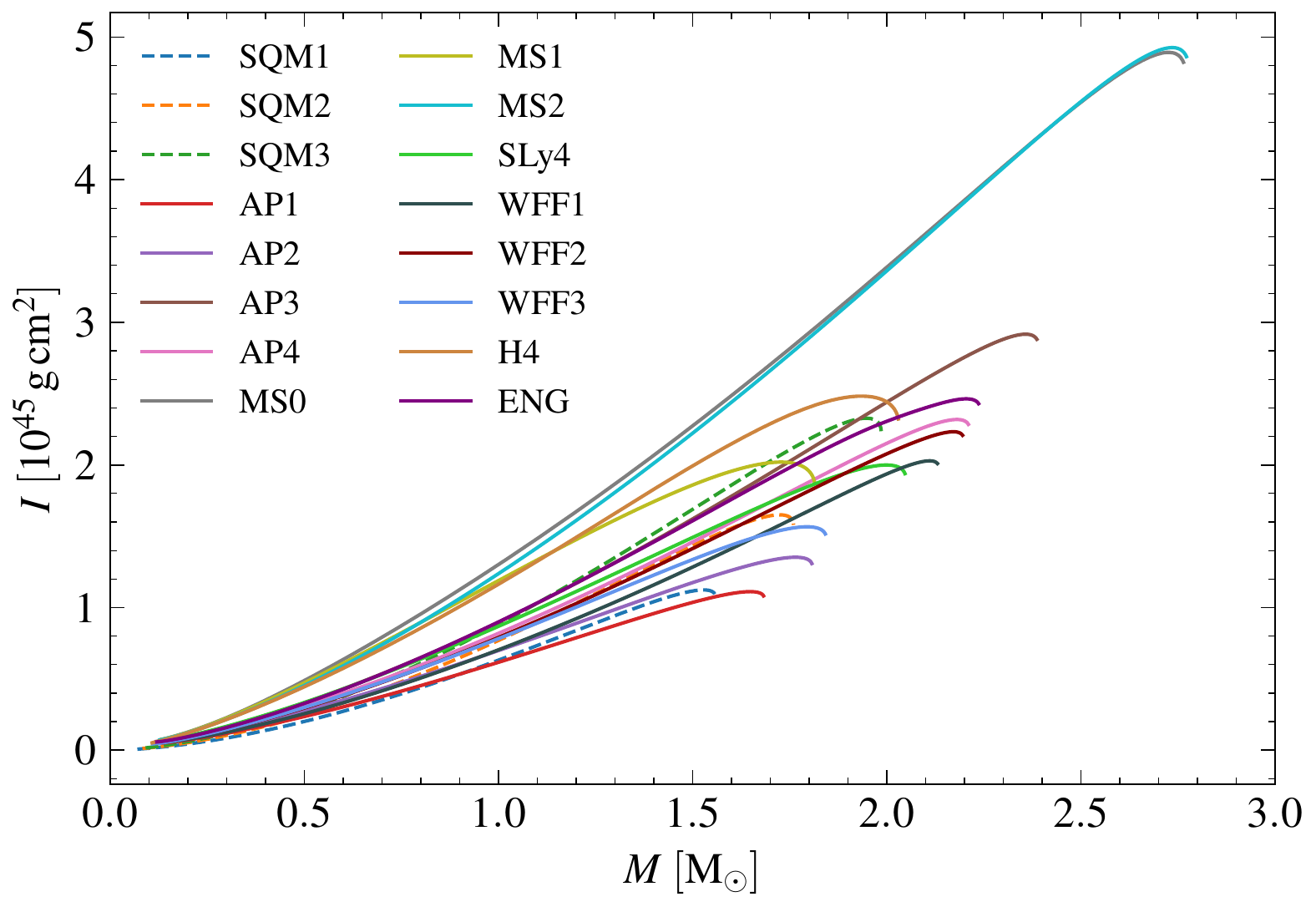}
	\vspace{-12pt}
	\caption{
		\label{fig: I_vs_M}
		The moment of inertia as a function of mass for different EOSs \citep{lattimerPrakash2001}. The SQM EOSs represent QS EOSs (dashed) and the others are NS EOSs (solid). One can see that for both QS and NS EOSs, the moment of inertia generally decreases for decreasing mass.
	}
\end{figure}

As for the change in moment of inertia of the magnetar, we calculated the moment of inertia for many different EOSs in Fig.~\ref{fig: I_vs_M}  following \cite{lattimerPrakash2001}. Except for extreme cases near the onset of instability, one finds that mass loss always results in a decrease in the moment of inertia for the magnetar, regardless of whether the magnetar has a NS-like or QS-like EOS (see also \cite{haenselZdunikShaefer1986}). This has a direct consequence for the spin of the (uniformly rotating) magnetar. During a glitch or antiglitch event, we want to conserve angular momentum $J = I \Omega$, i.e.~$\Delta J = 0$, which results in 
\begin{align}
\label{conservation_of_angular_momentum}
\frac{\Delta \nu}{\nu_0} = - \frac{\Delta I}{I_0}~,
\end{align}
where $\Delta \nu$ and $\Delta I$ are the changes in the system's spin frequency and the system's moment of inertia, respectively. By ``system'', we mean the magnetar plus the trapped ejecta, since it is assumed the ejecta is trapped in the magnetosphere which is strongly coupled and rigidly rotating with the magnetar. The subscript `0' refers to the value immediately before the glitch or antiglitch event.

In general, $\Delta I$ will comprise of two parts
\begin{align}
\Delta I = \Delta I_\mathrm{magnetar} + \Delta I_\mathrm{ejecta}~,
\end{align}
where $\Delta I_\mathrm{magnetar}$ is the change of moment of inertia of the magnetar, which we know to be negative for mass loss, and $\Delta I_\mathrm{ejecta}$ is the change in moment of inertia due to the ejecta being present (compared to the pre-glitch configuration, when $I_\mathrm{ejecta} = 0$). During the glitch/antiglitch, the ejecta moves radially outwards so $\Delta I_\mathrm{ejecta}$ will always be positive or zero, with maximal value when $\alpha = 90\text{\textdegree}$ and zero when $\alpha = 0\text{\textdegree}$.

The idea of glitches and antiglitches therefore comes quite naturally from this description. The decrease in the moment of inertia of the magnetar will want to increase the spin frequency of the system whereas the increase in the moment of inertia due to the trapped ejecta will want to decrease the spin frequency of the system. These two effects are in contest against one another and whichever effect dominates decides whether a glitch or antiglitch occurs.

Moreover, a time-varying non-axisymmetric mass distribution is established when the trapped ejecta rotates with the (spherical) magnetar. This is all that is required for GWs to be emitted \citep{thorne1980, andersson2019GWbook}. The trapped ejecta induces a biaxial mass distribution with more mass along the magnetic axis. In general, when a body does not rotate about one of its principal axes, its moment of inertia tensor is no longer diagonal. As a result, the moment of inertia tensor  in the inertial reference frame becomes time dependent so in order to conserve angular momentum, the angular frequency vector of the body must evolve. In other words, the body is set into free precession \citep{jonesAndersson2001}. 

The GW problem for freely precessing biaxial NSs has already been solved by \cite{zimmermannSzedenits1979}, \citet{cutlerJones2000} and \cite{jonesAndersson2002} (see \citet{gaoetal2020} for a treatment of triaxial NSs) but here, we extend some of their calculations on the assumption that the ejecta is responsible for the biaxial mass distribution, rather than an intrinsic deformation of the NS, often parametrised by the oblateness $\varepsilon$ or the difference in the moment of inertia between the deformed axis and one of the two other axes. In our model, we essentially map $\varepsilon$ onto a combination of $\delta M$ and $l$, the mass of the trapped ejecta and the ejecta height, which introduces one extra model parameter compared to the standard biaxial NS model. The GW radiation problem for our model will be addressed in more detail in Section~\ref{section_gravitational_wave_radiation}.

Finally, a word on the time-scales involved with our model. We have, by hand, assumed the ejected mass instantaneously appears above the magnetic poles at the moment of the glitch/antiglitch. This, of course, is unphysical and there must be some time-scale associated with this rapid ejection of matter from the magnetar. At least for NS glitches, it has been observed that this time-scale (the ``glitch rise time'') is very short, potentially shorter than 12.6~s \citep{ashtonetal2019}. In our model, when the particles are ejected, they move along open field lines so are accelerated to relativistic speeds \citep{rudermanSutherland1975} but shortly after, they must somehow decelerate to zero radial velocity. At least for the outbound journey, it is possible that the ejecta reaches its destination within an apparently instantaneous time-scale which gives justification for our modelling. The process of deceleration would depend on the details of the trapping mechanism, which we are agnostic to, but we expect the effect from higher order magnetic multipoles to be the most dominant factor. It is also expected that this deceleration would produce electromagnetic radiation which should be investigated further, especially on whether it could produce subsequent FRBs.

There is also the question of how long the ejecta remains trapped. We will argue this from an astrophysical viewpoint and the answer directly links with how long the GW signal is expected to last for. In our model, what determines how long the ejecta is trapped for is however long it takes for a pulsed radio emission to be seen after the glitch or antiglitch. Our reasoning for this is given in the following paragraph. If a pulsed radio emission is not seen, then our model is not constraining so introduces the trapping time-scale $T_\text{trap}$ as another model parameter. 

Usually, pulsed radio emission is thought to originate from the polar cap above the magnetic poles \citep[e.g.][]{lyneGraham-Smith2012, philippovTimokhinSpitkovsky2020}. The idea is that a strong electric field, induced by the rotating magnetic field, accelerates charged particles across a vacuum ``polar gap'' above the magnetic poles and at a certain height, the accelerated particles have enough energy to undergo electron-positron pair production. One of the particles from the pair gets further accelerated and produces another pair, leading to a pair production cascade. The acceleration of particles along curved magnetic field lines then leads to curvature radiation. Highly energetic curvature radiation photons are released but their propagation is interrupted by spontaneous pair production, decreasing the photon's energy until they reach radio energies and become unaffected by further pair production. These radio photons then stream outwards and are detected by our radio telescopes. Clearly, the polar gap is important for radio emission to occur. If this gap is filled by ejecta, then no gap/electric field exists and hence, radio emission is suppressed. Once the ejecta is no longer trapped, the polar gap re-establishes and pulsed radio emission is observed.

We know that there was a period of pulsed radio emission after the latest SGR J1935+2154 antiglitch \citep{younesetal2023}. The antiglitch occurred on 2020 October 5 ($\pm 1~\text{d}$) and the pulsed radio emission began from 2020 October 9 \citep{zhuetal2020, zhuetal2023}. This means that, in this example, we expect the ejecta to be trapped for around 4 days. This is how long we would expect the GWs to last for too.

\subsection{Glitch or antiglitch size}

We will now make the model more quantitative by introducing some simple equations. In the toy model, the moment of inertia of the system (magnetar plus ejecta) before the glitch/antiglitch is simply just the moment of inertia of the magnetar which we approximate using the uniform density approximation
\begin{align}
I_0 = \frac{2}{5} M_0 R_0^2~,
\end{align}
and once mass $\delta M > 0$ is ejected to a height $l$ above the pre-glitch radius $R_0$, the moment of inertia of the system becomes
\begin{align}
I_\text{after} = \frac{2}{5} (M_0-\delta M) (R_0 + \delta R)^2 + 2\left(\frac{1}{2}\delta M\right)(R_0+l)^2 \sin^2\alpha~,
\end{align}
where $\delta R$ is positive if the magnetar has a NS-like EOS and is negative for a QS-like EOS. In our model, we require $l > |\delta R|$. We define the change in moment of inertia as
\begin{align}
\Delta I \equiv	I_\text{after} - I_0~,
\end{align}
and when we calculate the fractional change in the moment of inertia, to first order in small quantities $\delta M \ll M_0$ and $\delta R \ll R_0$, we get
\begin{align}
\label{change_in_MoI_general}
\frac{\Delta I}{I_0} \approx 2 \left(\frac{\delta R}{R_0}\right) - \left(\frac{\delta M}{M_0}\right) + \frac{5}{2}\left(\frac{\delta M}{M_0}\right)\left(1+\frac{l}{R_0}\right)^2\sin^2\alpha~.
\end{align}
By equation~(\ref{conservation_of_angular_momentum}), one can see that the left hand side is just minus the fractional change in spin frequency. One can also see that the equation simplifies nicely if we convert the $\delta R/R_0$ term into one that is a multiple of $\delta M/M_0$ which we will do in the following subsections. As mentioned previously, the change in radius is different for NSs and QSs so we will deal with them separately.

\subsubsection{Quark stars}

\begin{figure}
	\includegraphics[width=\linewidth]{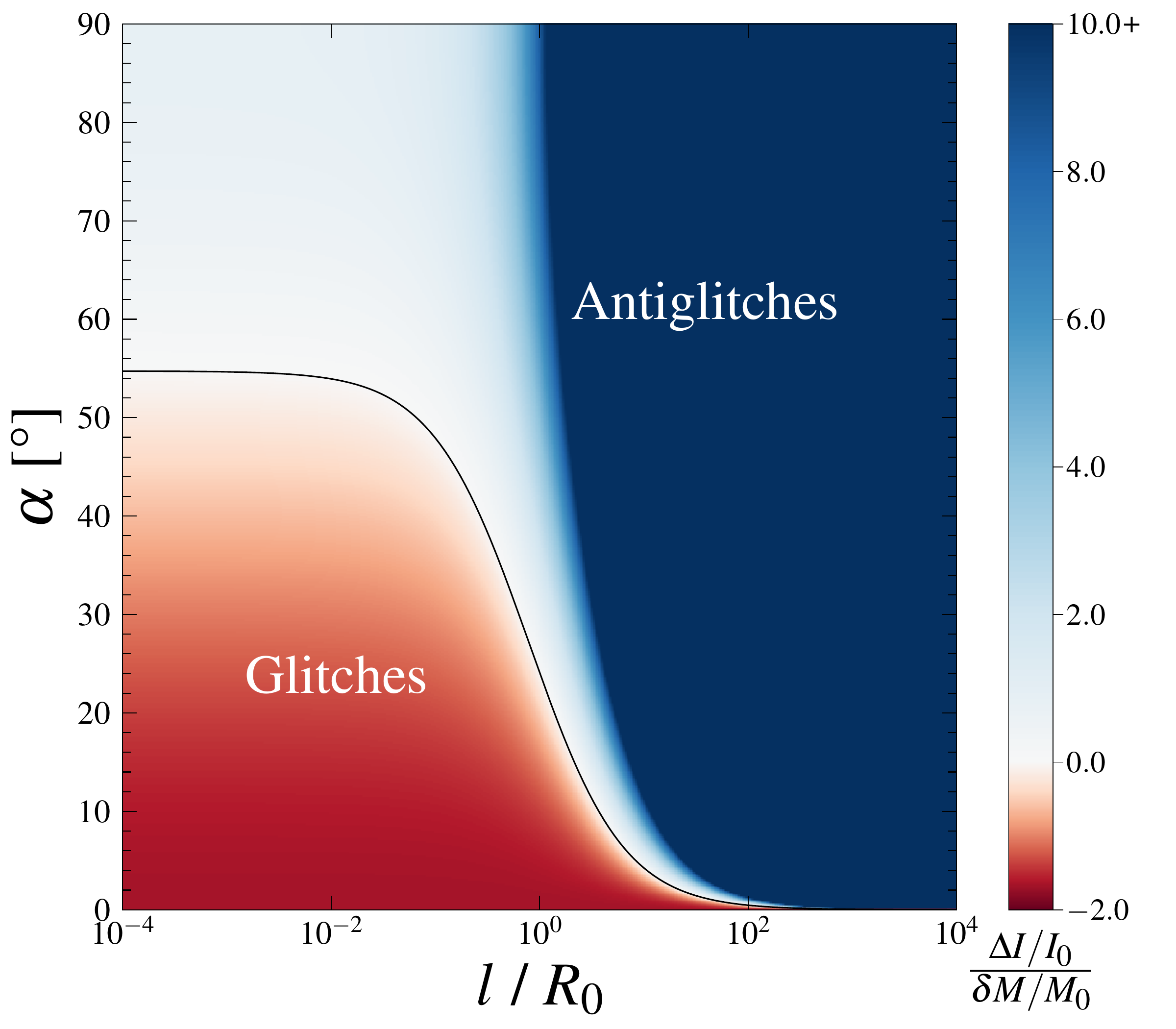}
	\vspace{-12pt}
	\caption{
		\label{fig: qs_glitch_antiglitch_parameter_space}
		A colour plot showing the value of $(\Delta I/I_0)/(\delta M/M_0)$ (or similarly, the square brackets in equation~(\ref{change_in_MoI_quark_star})) for QSs as a function of ejecta height $l$ (normalised by the magnetar's pre-glitch radius $R_0$) and the magnetic inclination angle $\alpha$. The red region represents the combinations of ($l/R_0$, $\alpha$) that permit glitches and the blue region is the same but for antiglitches. The black curve separates these two regions. Note that the ejecta height has a firm upper limit of the light cylinder radius, $R_\text{LC} = c / \Omega$, where co-rotating matter would be travelling at the speed of light $c$. For SGR J1935+2154, this occurs at $l/R_0 \sim 1.6 \times 10^{4}$. From equation~(\ref{corotation_radius}), this magnetar's co-rotation radius occurs at $l/R_0 \approx 370$.
	}
\end{figure}

We will first evaluate the case where the magnetar has a QS-like EOS since this is the easiest to understand. QSs act in a ``na\"ive'' sense, where decreasing the mass also decreases the radius \citep[e.g.][]{ozelFreire2016}. Some defining properties of QSs is that they are self-bound by the strong force and are nearly incompressible.
As a result, QSs, to a very good approximation, can be treated as having constant density \citep{haenselZdunikShaefer1986, alcockFarhiOlinto1986, gaoetal2022}. The mass lost from a thin spherical shell near the magnetar's surface can therefore be written as
\begin{align}
\label{delta_M_quark_star}
\delta M \approx - 4 \pi R_0^2 \bar{\rho} \delta R = - 3 \frac{M_0}{R_0}\delta R~,
\end{align}
where the average mass density $\bar{\rho}$ is given by
\begin{align}
\bar{\rho} = \frac{3 M_0}{4\pi R_0^3}~.
\end{align}
Rearranging equation~(\ref{delta_M_quark_star}), we get
\begin{align}
\frac{\delta R}{R_0} = - \frac{1}{3} \frac{\delta M}{M_0}~,
\end{align}
which, when substituted into equation~(\ref{change_in_MoI_general}), gives 
\begin{align}
\label{change_in_MoI_quark_star}
\frac{\Delta I}{I_0} \approx \left(\frac{\delta M}{M_0}\right) \left[ \frac{5}{2}\left(1+\frac{l}{R_0}\right)^2\sin^2\alpha - \frac{5}{3} \right]~.
\end{align}
From factorising $\delta M/M_0$, we end up with a much simpler equation. We know that $\delta M/M_0$ is positive, so ultimately, the sign of the square brackets determines the sign of $\Delta I/I_0$ and hence $\Delta \nu/\nu_0$ by equation~(\ref{conservation_of_angular_momentum}). The sign of the square brackets is determined by the choice of ejecta height $l$ and magnetic inclination angle $\alpha$. The value of the square brackets, or similarly $(\Delta I/I_0)/(\delta M/M_0)$, can be plotted as a function of the model's parameter space ($l/R_0$, $\alpha$), the results of which can be found in Fig.~\ref{fig: qs_glitch_antiglitch_parameter_space}. 

One can use this figure to calculate how much mass is required to be ejected for a given (observed) glitch size and for a given combination of ($l/R_0$, $\alpha$). For example, if we let $(l/R_0, \alpha)=(0.1, 40$\textdegree$)$, one finds $\delta M/M_0 \sim \Delta \nu/\nu_0$.

There are a few limits of equation~(\ref{change_in_MoI_quark_star}) that we can look at. Firstly, in the case where $\alpha \rightarrow 0\text{\textdegree}$, the first term vanishes and we are left with $\Delta I/I_0 = - (5/3) (\delta M/M_0)$ which guarantees a glitch. Physically, this limit represents the ejecta being held right above the rotational axis so does not contribute to any increase to the system's moment of inertia, but simultaneously, the magnetar loses mass meaning its contribution to the change in the system's moment of inertia is negative. The net result is a glitch.

There is also the limit where $\alpha \rightarrow 90\text{\textdegree}$ which leads to $\Delta I/I_0 \ge (5/6) (\delta M/M_0)$ and guarantees an antiglitch. The equals sign is taken when $l \rightarrow 0$ and $\Delta I/I_0$ gets correspondingly larger for larger $l$. This scenario corresponds to the ejecta being held at a position perpendicular to the rotation axis, so contributes the largest possible positive change to the system's moment of inertia for a given mass and ejecta height. It is so large that the decrease in the magnetar's moment of inertia is not enough to prevent an antiglitch.

One might ask whether there is a middle ground between these two extreme values of $\alpha$ where neither a glitch or antiglitch occurs. We can think of this as a critical value of $\alpha$ where changing from below to above it changes a glitch to an antiglitch (and vice versa). In other words, we get glitches when $\alpha < \alpha_\text{crit}$ and antiglitches when $\alpha > \alpha_\text{crit}$. $\alpha_\text{crit}$ can be found by simply equating the square brackets to zero and, after rearranging, we get
\begin{align}
\label{alpha_crit_quark_star}
\alpha_\text{crit} = \sin^{-1}\left( \sqrt{\frac{2}{3}}\left(1+\frac{l}{R_0}\right)^{-1} \right)~.
\end{align}
This has been plotted as a black curve in Fig.~\ref{fig: qs_glitch_antiglitch_parameter_space}. One can evaluate the limits of large and small $l$. For $l/R_0 \rightarrow \infty$, $\alpha_\text{crit} \rightarrow 0\text{\textdegree}$ meaning that there is no smaller value of $\alpha$ that can allow a glitch, i.e.~glitches are not possible when $l/R_0 \rightarrow \infty$. On the other hand, one can interpret this as glitches are always possible for sufficiently small $\alpha$. We will soon see that this conclusion also holds for NSs. For $l/R_0 \rightarrow 0$, we find $\alpha_\text{crit} = \sin^{-1}(\sqrt{2/3}) \approx 54.7\text{\textdegree}$. The information gained from these two limits suggests that if $\alpha$ were ever determined to be larger than $\alpha_\text{crit} \approx 54.7\text{\textdegree}$ through some independent method, e.g. pulse polarisation fitting \citep{radhakrishnanCooke1969}, then our model predicts that only antiglitches will occur given the magnetar has a QS-like EOS.

We can also invert equation~(\ref{alpha_crit_quark_star}) to find what the critical ejecta height $l_\text{crit}$ is for a given $\alpha$, for $0 < \alpha < \sin^{-1}(\sqrt{2/3})$. We find
\begin{align}
\label{l_crit_quark_star}
\frac{l_\text{crit}}{R_0} = \frac{\sqrt{2} - \sqrt{3}\sin\alpha}{\sqrt{3}\sin\alpha}~,
\end{align}
where $l < l_\text{crit}$ leads to glitches and $l > l_\text{crit}$ leads to antiglitches. This reiterates the result that ejecta trapped at a lower (higher) height is more likely to have a glitch (antiglitch). It also means that glitches are always possible for sufficiently small $l$.

\subsubsection{Neutron stars}

Typically, when a NS loses mass, its radius increases or approximately stays constant. The amount by which it changes depends on the selected EOS but in the absence of certainty, we take the option to define a new phenomenological dimensionless ``EOS parameter'' $\gamma$ as
\begin{align}
\label{def_gamma}
\frac{\delta R}{R_0} \equiv \gamma \frac{\delta M}{M_0}~,
\end{align}
where $\gamma \ge 0$ and neatly encodes our uncertainty of the NS EOS. Doing this allows us to continue tackling this problem analytically so that we at least get a qualitative understanding of our model. One finds that $\gamma$ defined in this way is the same as defining the mass-radius relation to be
\begin{align}
\label{def_gamma_mass_radius}
M \propto R^{-\frac{1}{\gamma}}~,
\end{align}
near the NS's true mass, where we have also been careful to stay consistent with our convention of mass loss having $\delta M > 0$ such that $dM = -\delta M$. Interestingly, if we generalise $\gamma$ to be either positive or negative, then we can model QSs as having $\gamma = -1/3$.

Additionally, we can determine $\gamma$ by multiplying the inverse of the mass-radius gradient with the compactness of the object
\begin{align}
 \gamma = - \mathcal{C} \left(\frac{dM}{dR}\right)^{-1}~,
\end{align}
where $\mathcal{C} = M_0/R_0$ is the compactness. From this, we can compare to realistic EOSs to find that for stiffer EOSs, we get smaller values of $\gamma$ and for softer EOSs, we get larger values of $\gamma$. A stiff EOS means the NS is not very compressible (so has a large internal sound speed) and vice versa for a soft EOS. One can also calculate $\gamma$ analytically for analytical EOSs such as polytropes, which we will do later.

Putting equation~(\ref{def_gamma}) into (\ref{change_in_MoI_general}), we get
\begin{align}
\label{change_in_MoI_neutron_star}
\frac{\Delta I}{I_0} \approx \left(\frac{\delta M}{M_0}\right) \left[ \frac{5}{2}\left(1+\frac{l}{R_0}\right)^2\sin^2\alpha + (2\gamma - 1) \right]~.
\end{align}
Again, the sign of the square brackets tells us whether a glitch or antiglitch occurs. Note that the first term is always positive and the second term is positive whenever $\gamma > 1/2$. Therefore, we predict that for $\gamma \ge 1/2$, a NS will only be able to exhibit antiglitches. For $0 \le \gamma < 1/2$, glitches are still possible. In the limit of $\gamma = 0$, we see that the second term in the square brackets goes to $-1$. This is not as negative as the QS case, where the second term is $-5/3$. This leads to the conclusion that QSs give larger glitches compared to NSs, given all else is the same.

\begin{figure}
	\includegraphics[width=\linewidth]{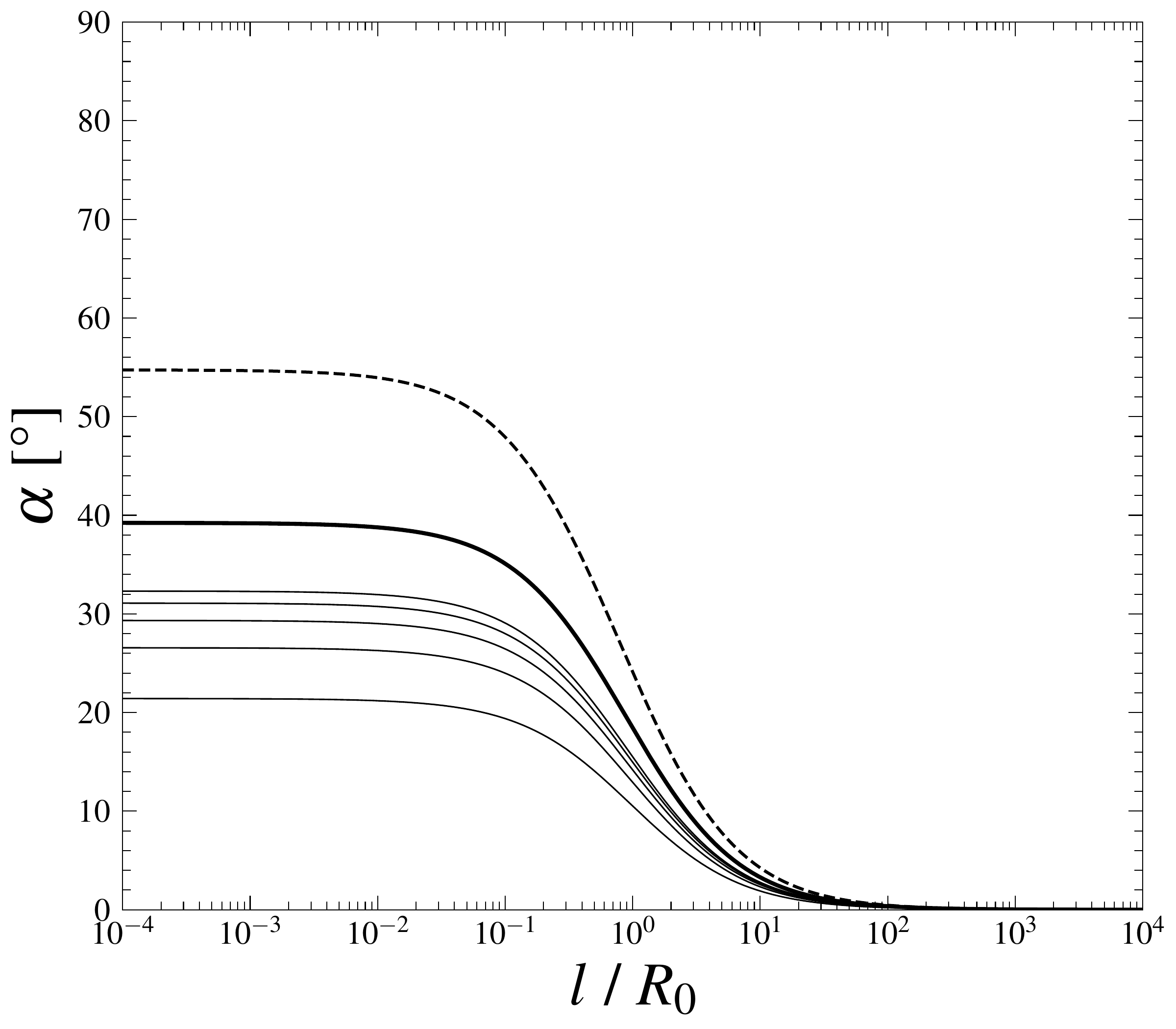}
	\vspace{-12pt}
	\caption{
		\label{fig: ns_glitch_antiglitch_parameter_space}
		A figure showing the critical boundary curves for different $\gamma$ values for NS-like EOSs (solid lines). Starting from the bottom curve and going up, the curves represent values of $\gamma = \frac{1}{3}, \frac{1}{4}, \frac{1}{5}, \frac{1}{6}, \frac{1}{7}$ until the bold curve, which is the limit of $\gamma = 0$. The dashed curve represents a QS-like EOS, which is also shown in Fig.~\ref{fig: qs_glitch_antiglitch_parameter_space}. If a selected combination of $l/R_0$ and $\alpha$ lies beneath a given curve, then a glitch will occur. If the selected point lies above a given curve, then an antiglitch will occur.
	}
\end{figure}

Assuming glitches are possible, i.e.~$0 \le \gamma < 1/2$, then the critical value of $\alpha$ that separates glitches and antiglitches is given by
\begin{align}
\label{alpha_crit_ns}
\alpha_\text{crit} = \sin^{-1}\left(\sqrt{\frac{2}{5} - \frac{4}{5}\gamma}\left(1+\frac{l}{R_0}\right)^{-1}\right)~,
\end{align}
where $\alpha < \alpha_\text{crit}$ yields glitches and $\alpha > \alpha_\text{crit}$ yields antiglitches. Like in the QS case, we can invert this to get the critical ejecta height 
\begin{align}
\label{l_crit_neutron_star}
\frac{l_\text{crit}}{R_0} = \frac{\sqrt{\frac{2}{5} - \frac{4}{5}\gamma} - \sin\alpha}{\sin\alpha}~.
\end{align}
In Fig.~\ref{fig: ns_glitch_antiglitch_parameter_space}, we plot the curve $\alpha_\text{crit}$ as a function of $l/R_0$, i.e. equation~(\ref{alpha_crit_ns}), for different values of $\gamma$. One finds that stiffer EOSs (lower values of $\gamma$) allow glitches in a larger fraction of the model's parameter space. Also, for $\alpha > \sin^{-1}(\sqrt{2/5}) \approx 39.2\text{\textdegree}$, no matter the value of $\gamma$, NSs are guaranteed to have an antiglitch according to our model.

At this point, we can make a meaningful comparison between the critical boundary curves of NSs and QSs. In Fig.~\ref{fig: ns_glitch_antiglitch_parameter_space}, the dashed curve represents the QS curve, which was derived in equation~(\ref{alpha_crit_quark_star}). We show in the figure that all NS curves, within the allowed $\gamma$ range, lie beneath the QS curve, meaning QSs have a larger parameter space in which they can glitch when compared to NSs. In fact, there is a region between the QS and (any of the) NS curves where only a QS would be able to explain a glitch, but not a NS. This means that if we were ever to observe a glitch from a magnetar, with known magnetic inclination angle $\sin^{-1}(\sqrt{2/5}) < \alpha < \sin^{-1}(\sqrt{2/3})$, or similarly $39.2\text{\textdegree} < \alpha < 54.7\text{\textdegree}$, then, according to our model, this must be a magnetar with a QS-like EOS.

In our model, we introduced $\gamma$ as a new parameter but if we use a polytropic EOS, we can link $\gamma$ to the familiar adiabatic index $\Gamma$ and polytropic index $n$ \citep[e.g.][]{shapiroTeukolsky1983}. A polytropic EOS is one where the pressure is a power law of the mass density, defined as
\begin{align}
P \equiv \kappa \rho^{\Gamma} \equiv \kappa \rho^{1 + \frac{1}{n}}~,
\end{align}
where $\kappa$ is a constant of proportionality and the second equality defines the polytropic index $n$ in terms of the adiabatic index $\Gamma = 1 + 1/n$. In Appendix~\ref{appendix_A}, we show the mass-radius relation for a polytropic EOS in Newtonian gravity with fixed central mass density is given by
\begin{align}
\label{mass_radius_relation_polytrope}
M \propto R^{\frac{3-n}{1-n}}~.
\end{align}
One can look at some interesting values of $n$. For $n = 3$, we find that $M$ becomes independent of $R$. On a mass-radius diagram, this corresponds to a horizontal line. For $n = 1$, we find $M \rightarrow \infty$, which is a vertical line on the mass-radius diagram. These two values of $n$ represent the two extremes for NSs and as we will see shortly, $n = 1$ corresponds to $\gamma = 0$ and $n = 3$ corresponds to $\gamma \rightarrow \infty$.

\begin{figure}
	\includegraphics[width=\linewidth]{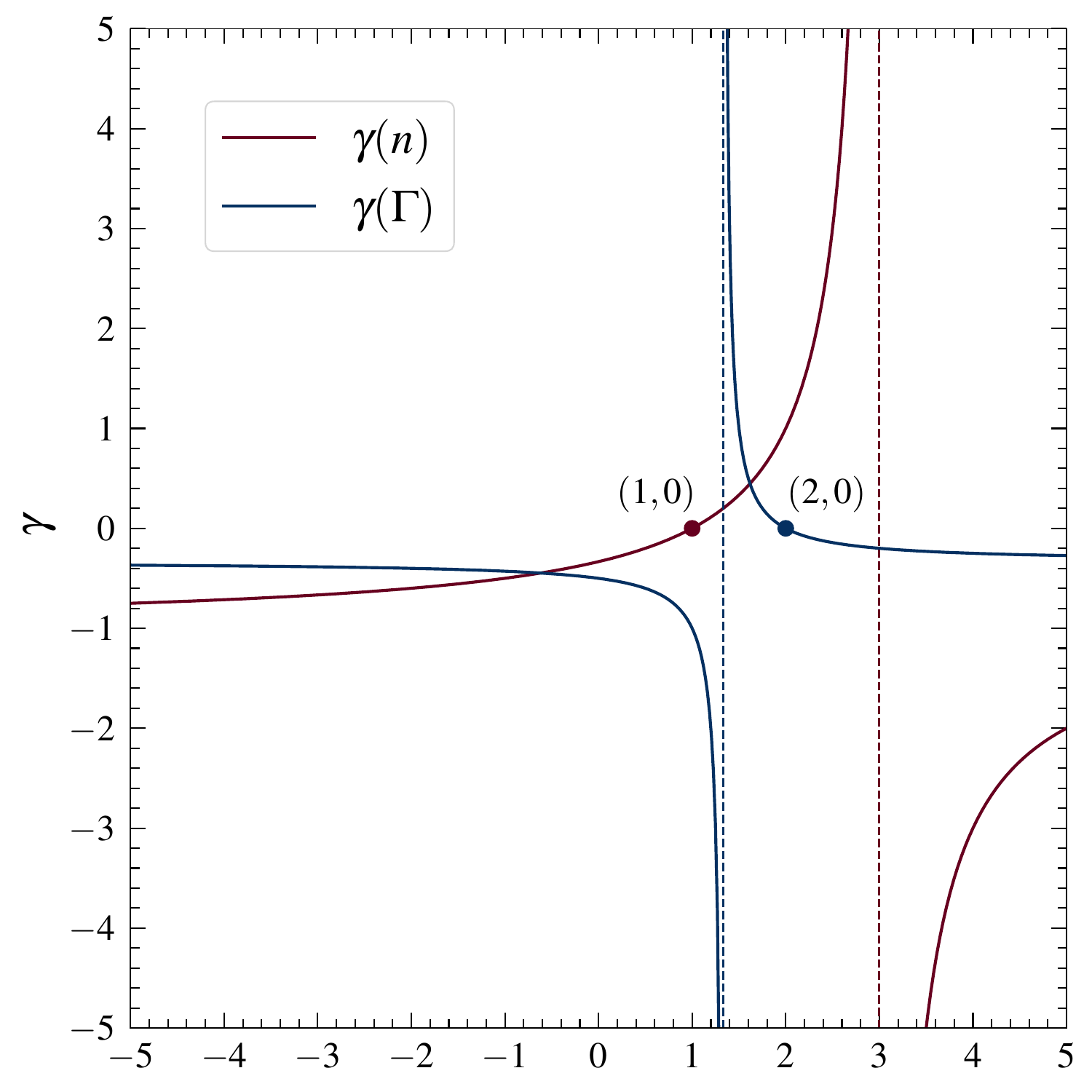}
	\vspace{-12pt}
	\caption{
		\label{fig: gamma_function_of_n_and_Gamma}
		A graph showing $\gamma$ as a function of the polytropic index $n$ (red) and the adiabatic index $\Gamma$ (blue). The asymptotes are at $\Gamma = \frac{4}{3}$ and $n = 3$. The highlighted points on the curves represent values where $\gamma = 0$.
	}
\end{figure}

To get $\gamma$ as a function of $n$, we directly compare equations~(\ref{def_gamma_mass_radius}) and (\ref{mass_radius_relation_polytrope}) to find
\begin{align}
\label{gamma_function_n_and_Gamma}
\gamma = \frac{n - 1}{3 - n} = \frac{2 - \Gamma}{3 \Gamma - 4}~,
\end{align}
where $1 \le n < 3$ and $\frac{4}{3} < \Gamma \le 2$ if $\gamma \ge 0$. Fig.~\ref{fig: gamma_function_of_n_and_Gamma} shows $\gamma$ as a function of $n$ and also as a function of $\Gamma$. Note that \cite{readetal2009} determined $\Gamma$ for many EOSs using a piecewise polytrope description, where they separated the NS into three regions according to mass density, each with a corresponding value of $\Gamma$. The regions began from the bottom of the crust and extended inwards towards the centre. They found that in all three sections $\Gamma$ varied from around 2 to 4 for hadronic EOSs. For our $\gamma$ parameter, this corresponds to $-\frac{1}{4} < \gamma < 0$ which is out of our validity range. This inconsistency is due to the fact that the equation relating $\Gamma$ to $\gamma$, equation~(\ref{gamma_function_n_and_Gamma}), was derived in Appendix~\ref{appendix_A} using Newtonian gravity instead of general relativity. The effects of gravity are stronger in general relativity resulting in higher values for the allowed range of $\Gamma$ \citep{tooper1964, readetal2009}. At the same time, the behaviour of the piecewise polytrope EOS is still NS-like, hence we would still expect $\gamma > 0$ in the full treatment using general relativity. 

If the mass lost were solely from the crust, then a crustal EOS should be used. A typical crustal EOS is the Sly EOS \citep{douchinHaensel2001} though there are numerous other crustal EOSs available too \citep{rusteretal2006, chamelHaensel2008}. \cite{rusteretal2006} calculated $\Gamma$ for many crustal EOSs, below nuclear drip density $\rho \approx 4 \times 10^{11}~\text{g~cm}^{-3}$, and found that they all had $\frac{4}{3} < \Gamma < \frac{9}{5}$, which is then consistent with our validity range. 

\section{Gravitational wave radiation}
\label{section_gravitational_wave_radiation}

We now move onto the GW aspect of our toy model. As mentioned in Section~\ref{subsection_the_trapped_ejecta_model}, when the ejecta becomes trapped above the magnetic poles, the moment of inertia tensor of the system changes so the system must undergo free precession in order to conserve angular momentum \citep{jonesAndersson2001}. The trapped ejecta deforms the system into a biaxial configuration and this has been shown to emit GWs \citep{zimmermannSzedenits1979, cutlerJones2000, jonesAndersson2002}. 

We begin with Section~\ref{subsection_precession_biaxial_deformed_bodies} where we summarise the basic equations for a precessing biaxial body, parametrised by the difference in moment of inertia due to some deformation $\Delta I_\text{d}$ and ``wobble angle'' $\alpha$ (so we can draw similarities with the trapped ejecta model). Then, in Section~\ref{subsection_gravitational_waves_from_trapped_ejecta}, we generalise the standard biaxial precession equations for the trapped ejecta system which will be parametrised by $\delta M$, $l$ and $\alpha$. Here, we will state the GW strain, GW luminosity, GW torque and the total GW energy emitted for a given trapping time-scale $T_\text{trap}$. Finally, in Section~\ref{subsection_SNR_detectability}, we calculate the signal-to-noise ratio (SNR) achievable with future GW detectors.

\subsection{Precession of biaxially deformed bodies}
\label{subsection_precession_biaxial_deformed_bodies}

In the precession problem, it is useful to define a body reference frame (indicated by $i, j, ... = 1, 2, ...$ indices) and an inertial reference frame (indicated by $a, b, ... = xx, xy, ...$ indices). The body frame can be used to exploit the symmetry of the biaxial system but we must eventually change into an inertial frame in order to evaluate the GW radiation. This is done by applying transformations in the form of rotations.

Firstly though, we define the principal axes of the biaxial body in the body frame as $I_1$, $I_2 (=I_1)$ and $I_3$ and assume the deformation acts along the 3-axis, giving rise to a difference in moment of inertia
\begin{align}
\Delta I_\text{d} \equiv I_3 - I_1~,
\end{align}
where $\Delta I_\text{d} > 0$ for an oblate system and $\Delta I_\text{d} < 0$ for a prolate system (which is what we will be mainly concerned with in the trapped ejecta model). $\Delta I_\text{d}$ is linked to the more commonly-known poloidal ellipticity, which is defined as
\begin{align}
\varepsilon \equiv \frac{I_3 - I_1}{I_1} = \frac{\Delta I_\text{d}}{I_1}~,
\end{align}
which can be used in all subsequent equations if one wanted to re-express the equations in terms of the poloidal ellipticity.

In such a deformed system, the moment of inertia tensor in the body frame is 
\begin{align}
\label{moment_of_inertia_biaxial}
I_{ij} = 
\begin{bmatrix}
I_1 & 0 & 0 \\
0 & I_1 & 0 \\
0 & 0 & I_3 \\
\end{bmatrix}
~,
\end{align}
where, in general, the moment of inertia tensor is defined as
\begin{align}
I_{ij} \equiv \int_{V} \rho (r^2 \delta_{ij} - x_i x_j) dV~,
\end{align}
where $\delta_{ij}$ is the Kronecker delta, $r^2 = x_1^2 + x_2^2 + x_3^2$ and $x_i$ is the position vector in a Cartesian basis. For GW calculations, it is also important to define the mass quadrupole moment tensor 
\begin{align}
\mathcal{I}_{ij} \equiv \int_{V} \rho x_i x_j dV~,
\end{align}
as well as the trace-reduced mass quadrupole moment tensor
\begin{align}
\mathsout{\mathcal{I}}_{ij} \equiv \int_{V} \rho (x_i x_j - \frac{1}{3} \delta_{ij} r^2)dV = \mathcal{I}_{ij} - \frac{1}{3}\delta_{ij} \text{Tr}(\mathcal{I})~,
\end{align}
which is traceless by definition. One can quickly see from the definition of the mass quadrupole moment that for a biaxial system, we will have $\mathcal{I}_{ij} = \text{Diag}[\mathcal{I}_1, \mathcal{I}_1, \mathcal{I}_3]$, similar in form to equation~(\ref{moment_of_inertia_biaxial}). This is not surprising given the moment of inertia tensor can be written as
\begin{align}
I_{ij} = \delta_{ij} \text{Tr}(\mathcal{I}) - \mathcal{I}_{ij}~,
\end{align}
meaning the difference between any of the diagonal components differs only by a minus sign
\begin{align}
I_{ii} - I_{jj} = - (\mathcal{I}_{ii} - \mathcal{I}_{jj})~,
\end{align}
where repeated indices in this equation are not summed over.

At this stage, it is important to transform from the body frame to an inertial frame. For now, we select an inertial frame that is located far above the angular momentum $J$-axis, i.e.~the viewing inclination angle is $\iota = 0\text{\textdegree}$. By construction, the biaxial body is inclined from the $J$-axis (which is also the inertial $z$-axis in this case) by angle $\alpha$ and is rotating about the $J$-axis at rate $\dot{\phi}= J/I_1$. Note that $\dot{\phi}$ differs slightly from $\Omega$, the electromagnetically observed angular frequency, by 
\begin{align}
\label{difference_in_frequencies}
\dot{\phi} - \Omega = \frac{\Delta I_\text{d}}{I_1} \Omega \left(1 + \mathcal{O}(\alpha^2)\right)~,
\end{align}
which is approximately equal to the (body frame) precessional frequency of the system and comes from carefully calculating the precessional behaviour \citep{zimmermannSzedenits1979, cutlerJones2000}.  

To get from the body frame to the inertial frame, we first enforce an active rotation about the $y$-axis (or $x$-axis) by $\alpha$ followed by an active rotation about the $z$-axis by $\dot{\phi} t + \phi_0$, where $\phi_0$ is the initial phase of the rotation. Collectively, this results in a transformation matrix of 
\begin{align}
\label{transformation_matrix}
R_{ai} = 
\begin{bmatrix}
\cos\alpha \cos \left(\dot{\phi}t +\phi _0\right) & -\sin \left(\dot{\phi} t +\phi _0\right) & \sin \alpha  \cos \left(\dot{\phi} t +\phi _0\right) \\
\cos \alpha  \sin \left(\dot{\phi} t +\phi _0\right) & \cos \left(\dot{\phi} t +\phi _0\right) & \sin \alpha  \sin \left(\dot{\phi} t +\phi _0\right) \\
-\sin \alpha  & 0 & \cos \alpha  \\
\end{bmatrix}
~.
\end{align}
One then simply applies the following transformation
\begin{align}
\label{transformation}
\mathsout{\mathcal{I}}^\text{(inertial)} = R^\text{T} \mathsout{\mathcal{I}}^\text{(body)} R ~,
\end{align}
which preserves the trace of $\mathsout{\mathcal{I}}$. Now that $\mathsout{\mathcal{I}}_{ab}$ is in the inertial frame, we can deduce numerous GW quantities including the GW luminosity $\dot{E}_\text{GW}$, GW torque $\dot{J}_\text{GW}$ and GW strain $h$. 
The GW luminosity is calculated by
\begin{align}
\label{GW_luminosity}
\dot{E}_\text{GW} = \frac{1}{5} \frac{G}{c^5} \left\langle \mathsout{\dddot{\mathcal{I}}}_{ab} \mathsout{\dddot{\mathcal{I}}}^{\hspace{1pt}ab} \right\rangle~,
\end{align}
where the dots represent time derivatives in the inertial frame and the angled brackets represent an average over many periods. $\dot{E}_\text{GW} > 0$ corresponds to GW energy being lost from the system which gets radiated out to infinity. The GW torque is calculated by
\begin{align}
\label{GW_torque}
\dot{J}^a_\text{GW} = \frac{2}{5} \frac{G}{c^5} \varepsilon^{abc} \left\langle \mathsout{\ddot{\mathcal{I}}}^{\hspace{1pt}d}_{~~b} \mathsout{\dddot{\mathcal{I}}}_{cd} \right\rangle~,
\end{align}
where $\varepsilon^{abc}$ is the antisymmetric Levi-Civita symbol. Similarly, $\dot{J}^a_\text{GW} > 0$ corresponds to positive angular momentum being carried away from the system. Note that both the above calculations require time derivatives of the trace-reduced mass quadrupole moment. However, if we recall that the trace of the mass quadrupole moment is constant ($\text{Tr}(\mathcal{I}) = 2 \mathcal{I}_1 + \mathcal{I}_3$), then the time derivative of this will be no different from the time derivative of the trace-reduced mass quadrupole moment, i.e.~$\dot{\mathcal{I}}_{ab} = \mathsout{\dot{\mathcal{I}}}_{ab}$, so $\ddot{\mathcal{I}}_{ab}$ and $\dddot{\mathcal{I}}_{ab}$ can be used in equations~(\ref{GW_luminosity}) and (\ref{GW_torque}) if one so wished.

The GW strain is a little more difficult to obtain as it will depend on the direction of the observer, indicated by some viewing inclination angle $\iota$, defined as the angle between the source's angular momentum vector $J_i$ and the direction towards the observer, assuming the source is at the origin of some inertial frame. 

Moreover, it is important to ensure quantities such as $\mathsout{\mathcal{I}}_{ab}$ are calculated in the transverse-traceless gauge, indicated by superscript ``TT''  \citep[e.g.][]{andersson2019GWbook}. This is achieved by a series of projections and leads to the expression for the GW strain tensor
\begin{align}
\label{GW_strain_tensor}
h_{ab}^\text{TT} = \frac{2}{d} \frac{G}{c^4} \mathsout{\ddot{\mathcal{I}}}_{ab}^\text{TT} ~,
\end{align}
where $d$ is the distance to the source. In this gauge, the GW strain in the plus and cross polarisations can be read off directly from the GW strain tensor as
\begin{align}
h_{+} &= h_{xx}^\text{TT} = - h_{yy}^\text{TT}~, \\
h_{\times} &= h_{xy}^\text{TT} = h_{yx}^\text{TT}~.
\end{align}	
This calculation for biaxial sources was first done by \cite{zimmermannSzedenits1979} and later confirmed by \cite{jones2010}. The GW polarisations were found to be
\begin{align}
\label{h_plus}
h_{+} = \frac{2}{d}\frac{G}{c^4} \dot{\phi}^2 \Delta I_\text{d} \sin\alpha &\left[(1 + \cos^2\iota)\sin\alpha \cos (2(\dot{\phi} t + \phi_0) )\right. \nonumber \\
&~~~\left.+ \sin\iota\cos\iota \cos\alpha \cos (\dot{\phi} t+ \phi_0)  \right]~, \\
\label{h_cross}
h_{\times} = \frac{2}{d}\frac{G}{c^4} \dot{\phi}^2 \Delta I_\text{d} \sin\alpha &\left[2\cos\iota\sin\alpha \sin (2(\dot{\phi} t + \phi_0) )\right. \nonumber \\
&~~~\left.+ \sin\iota \cos\alpha \sin (\dot{\phi} t+ \phi_0)  \right]~.
\end{align}
One can see that there is GW emission at two frequencies, at $\dot{\phi}$ and $2\dot{\phi}$, and by equation~(\ref{difference_in_frequencies}), these are not quite the same as $\Omega$ and $2\Omega$. For a prolate body, the GW frequencies will be lower than the respective electromagnetically observed frequencies, i.e. $\dot{\phi} < \Omega$, but only by a small amount since $\Delta I_\text{d} \ll I_1$.

For an observer directly above the angular momentum axis, we have $\iota = 0\text{\textdegree}$ and we no longer get any $\dot{\phi}$ radiation. The GWs become exactly circularly polarised which motivates the definition
\begin{align}
h_{\text{max}, 2\dot{\phi}} \equiv \frac{4}{d}\frac{G}{c^4} \dot{\phi}^2 |\Delta I_\text{d}| \sin^2\alpha~,
\end{align}
which we call the maximum (single polarisation) GW amplitude of the $2\dot{\phi}$ radiation. Similarly, the maximal (single polarisation) GW amplitude for the $\dot{\phi}$ radiation occurs when $\iota = 90\text{\textdegree}$ which is when there is minimal $2\dot{\phi}$ radiation and no $\dot{\phi}$ radiation in the $h_{+}$ polarisation, and no $2\dot{\phi}$ radiation and maximal $\dot{\phi}$ radiation in the $h_{\times}$ polarisation, given by
\begin{align}
h_{\text{max}, \dot{\phi}} \equiv \frac{2}{d}\frac{G}{c^4} \dot{\phi}^2 |\Delta I_\text{d}| \sin\alpha\cos\alpha~.
\end{align}
We can take the ratio of the two to learn which radiation dominates for a given $\alpha$
\begin{align}
\frac{h_{\text{max}, 2\dot{\phi}}}{h_{\text{max}, \dot{\phi}}} = 2 \tan\alpha~,
\end{align}
 which shows $\dot{\phi}$ radiation dominates for small $\alpha$ and $2\dot{\phi}$ radiation dominates for large $\alpha$.

Next, we will write down the expressions for $\dot{E}_\text{GW}$ and $\dot{J}_\text{GW}$. They were also calculated by \cite{zimmermannSzedenits1979} and \cite{cutlerJones2000} and are
\begin{align}
\label{GW_luminosity_generic_case}
\dot{E}_\text{GW} =\frac{2}{5}\frac{G}{c^5} \dot{\phi}^6 (\Delta I_\text{d})^2 \sin^2\alpha(\cos^2\alpha + 16\sin^2\alpha)~, \\
\label{GW_torque_generic_case}
\dot{J}_\text{GW} = \frac{2}{5}\frac{G}{c^5} \dot{\phi}^5 (\Delta I_\text{d})^2 \sin^2\alpha(\cos^2\alpha + 16\sin^2\alpha)~,
\end{align}
which recovers the well-known relation
\begin{align}
\dot{E}_\text{GW} = \dot{\phi} \dot{J}_\text{GW}~.
\end{align}
Note that the $\cos^2\alpha$ term comes from the $\dot{\phi}$ radiation and the $16\sin^2\alpha$ term comes from the $2\dot{\phi}$ radiation. This means for sufficiently small $\alpha$, it is possible that the $\dot{\phi}$ radiation is stronger than the $2\dot{\phi}$ radiation, motivating GW searches at both $\approx\Omega$ and $\approx2\Omega$ frequencies \citep{LVK2022twoharmonics}. It is also worth noting that for both oblate and prolate sources, $\dot{E}_\text{GW}$ and $\dot{J}_\text{GW}$ are both positive, meaning both energy and positive angular momentum are being radiated from the source.

To get an estimate of how much GW energy has been radiated away, we define 
\begin{align}
\label{E_GW_uniform}
E_\text{GW} \equiv \dot{E}_\text{GW} T_\text{GW}~,
\end{align}
where $T_\text{GW}$ is the duration for which the GWs are emitted for. It could be limited by the duration of the precession (like in the trapped ejecta model) or by the maximum observation time from a GW detector, whichever is shorter. Note that equation~(\ref{E_GW_uniform}) is exact if the emitted GW has a constant amplitude. In any other case, the right hand side becomes the time integral of the GW luminosity.

Finally, a word on how GW radiation causes a back-reaction on the system. Of course, with the loss of positive angular momentum from the system, $\dot{\phi}$ will decrease. However, it was also discovered by \cite{cutlerJones2000} that the angle $\alpha$ will decrease for both oblate and prolate bodies. In other words, GW radiation causes alignment ultimately leading to aligned rotators. They provided time-scales for how long each of these effects take which, for our values of interest, are at least 
\begin{align}
\tau_{\dot{\phi}} > 1.2 \times 10^{15}~\text{yr}~\left(\frac{\nu}{1~\text{Hz}}\right)^{-4}\left(\frac{\Delta I_\text{d}/I_1}{10^{-6}}\right)^{-2}\left(\frac{I_1}{10^{45}~\text{g~cm}^2}\right)^{-1}~, \\
\tau_{\alpha} > 4.3 \times 10^{15}~\text{yr}~\left(\frac{\nu}{1~\text{Hz}}\right)^{-4}\left(\frac{\Delta I_\text{d}/I_1}{10^{-6}}\right)^{-2}\left(\frac{I_1}{10^{45}~\text{g~cm}^2}\right)^{-1}~.
\end{align}
Clearly, for our trapped ejecta model, we do not need to worry about the GW back-reaction though for faster and more deformed systems, such as millisecond magnetars \citep[e.g.][]{landerjones2020}, the back-reaction may play an important role.

\subsection{Gravitational waves from trapped ejecta model}
\label{subsection_gravitational_waves_from_trapped_ejecta}

Now that the basic equations have been introduced, this subsection should be fairly self-explanatory. To remind ourselves, in the trapped ejecta model, we have two point masses $\frac{1}{2}\delta M$, diametrically opposite from one another, being held at height $l$ above the pre-glitch radius $R_0$. The angle between the rotational axis and the axis connecting the ejecta is $\alpha$.

To begin, we write down the mass quadrupole tensor in the body frame
\begin{align}
\mathcal{I}_{ij} = 
\begin{bmatrix}
\mathcal{I}_\text{sph} & 0 & 0 \\
0 & \mathcal{I}_\text{sph} & 0 \\
0 & 0 & \mathcal{I}_\text{sph} + \delta M (l + R_0)^2 \\
\end{bmatrix}
~,
\end{align}
where we aligned the body frame 3-axis with the line connecting the two point masses. $\mathcal{I}_\text{sph}$ is the mass quadrupole for a spherical object which is a constant. Reducing the quadrupole moment tensor by its trace effectively removes all spherically symmetric components, so we do not need to worry about the exact value of $\mathcal{I}_\text{sph}$.

Then, we reduce by the trace and transform to the inertial frame, the details of which were mentioned in equation~(\ref{transformation}). After taking an appropriate number of time derivatives, we can get $\dot{E}_\text{GW}$ and $\dot{J}_\text{GW}$, which are
\begin{align}
\label{GW_luminosity_trapped_ejecta}
\dot{E}_\text{GW} =\frac{2}{5}\frac{G}{c^5} \dot{\phi}^6 (\delta M)^2 \left(R_0 + l\right)^4 \sin^2\alpha(\cos^2\alpha + 16\sin^2\alpha)~, \\
\dot{J}_\text{GW} =\frac{2}{5}\frac{G}{c^5} \dot{\phi}^5 (\delta M)^2 \left(R_0 + l\right)^4 \sin^2\alpha(\cos^2\alpha + 16\sin^2\alpha)~.
\end{align}
In the limit of $l = 0$, equation~(\ref{GW_luminosity_trapped_ejecta}) is consistent with equation~(7) of \cite{sousaCoelhadeAraujo2020}, who were primarily concerned about GWs from accreting white dwarfs. Comparing the above equations to the generic case in equations~(\ref{GW_luminosity_generic_case}) and (\ref{GW_torque_generic_case}), we see that in the trapped ejecta model
\begin{align}
\Delta I_\text{d} = - \delta M \left(R_0 + l\right)^2 ~,
\end{align}
as expected. This can be also written in terms of the poloidal ellipticity as
\begin{align}
\varepsilon = - \frac{\delta M \left(R_0 + l\right)^2}{I_1}~.
\end{align}
A minus sign was added to $\Delta I_\text{d}$ since we expect $\Delta I_\text{d} < 0$ for the prolate configuration in the trapped ejecta model. The degeneracy between the ejecta mass and ejecta height means that GW observations alone will not be able to disentangle $\delta M$ and $l$. Small $\delta M$ at large $l$ can give the same GW luminosity/torque as large $\delta M$ at small $l$. Using the above expression for $\Delta I_\text{d}$, it is now straightforward to map the generic case onto the trapped ejecta case, which leads to a GW energy of 
\begin{align}
E_\text{GW} =\frac{2}{5}\frac{G}{c^5} \dot{\phi}^6 (\delta M)^2 \left(R_0 + l\right)^4 T_\text{GW} \sin^2\alpha(\cos^2\alpha + 16\sin^2\alpha)~,
\end{align}
and maximal (single polarisation) GW amplitudes of
\begin{align}
h_{\text{max}, 2\dot{\phi}} &\equiv \frac{4}{d}\frac{G}{c^4} \dot{\phi}^2 \delta M \left(R_0 + l\right)^2 \sin^2\alpha~, \\
h_{\text{max}, \dot{\phi}} &\equiv \frac{2}{d}\frac{G}{c^4} \dot{\phi}^2 \delta M \left(R_0 + l\right)^2 \sin\alpha\cos\alpha~.
\end{align}

Finally, we make the link to glitches and antiglitches using the equations derived in Section~\ref{section_a_toy_model_for_glitches_and_antiglitches}. Instead of having $\delta M$ in the equations above, we can make the equations depend on the observable $\Delta \nu / \nu_0$ using equations~(\ref{change_in_MoI_neutron_star}) and (\ref{conservation_of_angular_momentum}). This results in
\begin{align}
\Delta I_\text{d} = M_0 R_0^2 \left(\frac{\Delta \nu}{\nu_0}\right) \mathcal{G}~,
\end{align}
where $\mathcal{G}$ is a factor that is determined by the geometry of the magnetar system as well as the magnetar's EOS and is defined as
\begin{align}
\label{def_G}
\mathcal{G} \equiv \frac{\left(1 + \frac{l}{R_0}\right)^2}{\frac{5}{2}\left(1 + \frac{l}{R_0}\right)^2 \sin^2\alpha + (2\gamma - 1)}~.
\end{align}
In the trapped ejecta model, we require $\Delta I_\text{d} < 0$. This is satisfied by both positive and negative $\Delta \nu / \nu_0$ so long as $\mathcal{G}$ has the opposite sign to $\Delta \nu / \nu_0$. This is done by an appropriate choice of $l$ (say) such that $l < l_\text{crit}$ for glitches and $l > l_\text{crit}$ for antiglitches, as detailed in equation~(\ref{l_crit_neutron_star}). Another choice is to fix $l$ and choose $\alpha < \alpha_\text{crit}$ for glitches or $\alpha > \alpha_\text{crit}$ for antiglitches, as detailed in equation~(\ref{alpha_crit_ns}).

Similarly, the GW energy can be written as
\begin{align}
E_\text{GW} =\frac{2}{5}\frac{G}{c^5} \dot{\phi}^6 M_0^2 R_0^4 T_\text{GW} \left(\frac{\Delta \nu}{\nu_0}\right)^2 \mathcal{G}^2 \sin^2\alpha(\cos^2\alpha + 16\sin^2\alpha)~,
\end{align}
and the GW luminosity and torque as
\begin{align}
\dot{E}_\text{GW} =\frac{2}{5}\frac{G}{c^5} \dot{\phi}^6 M_0^2 R_0^4 \left(\frac{\Delta \nu}{\nu_0}\right)^2 \mathcal{G}^2 \sin^2\alpha(\cos^2\alpha + 16\sin^2\alpha)~, \\
\dot{J}_\text{GW} =\frac{2}{5}\frac{G}{c^5} \dot{\phi}^5 M_0^2 R_0^4 \left(\frac{\Delta \nu}{\nu_0}\right)^2 \mathcal{G}^2 \sin^2\alpha(\cos^2\alpha + 16\sin^2\alpha)~.
\end{align}
One should note that both $\dot{E}_\text{GW}$ and $\dot{J}_\text{GW}$ are positive definite for both glitches and antiglitches. Therefore, the trapped ejecta model predicts an increase in the spin-down rate after a glitch or antiglitch, i.e.~$\Delta \dot{\nu} < 0$ or $\Delta \dot{\nu} / \dot{\nu}_0 > 0$, for a duration equal to the time the ejecta is trapped for.

Lastly, the maximum (single polarisation) GW amplitudes for the trapped ejecta model are
\begin{align}
h_{\text{max}, 2\dot{\phi}} &= - \frac{4}{d}\frac{G}{c^4} \dot{\phi}^2 M_0 R_0^2 \left(\frac{\Delta \nu}{\nu_0}\right) \mathcal{G} \sin^2\alpha~, \\
h_{\text{max}, \dot{\phi}} &= - \frac{2}{d}\frac{G}{c^4} \dot{\phi}^2 M_0 R_0^2 \left(\frac{\Delta \nu}{\nu_0}\right) \mathcal{G} \sin\alpha\cos\alpha~.
\end{align}

\subsection{Signal-to-noise ratio}
\label{subsection_SNR_detectability}

\begin{figure}
	\includegraphics[width=\linewidth]{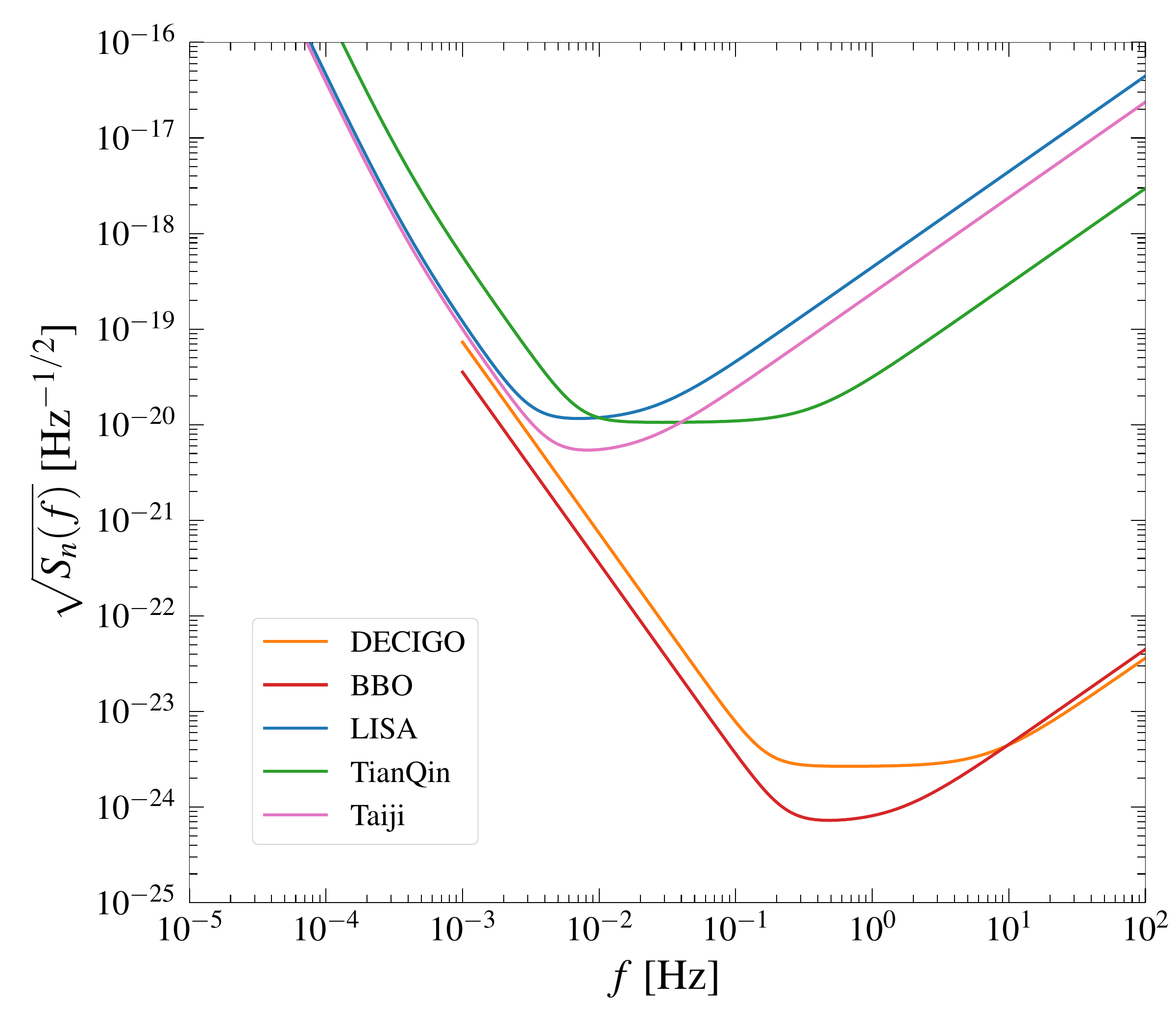}
	\vspace{-12pt}
	\caption{
		\label{fig: space_based_sensitivity_curves}
		The sensitivity curves (amplitude spectral density) $\sqrt{S_\text{n}}$ of future space-based detectors as a function of GW frequency $f$. Decihertz detectors include DECIGO and Big Bang Observer (BBO) and their sensitivity curves were obtained from 
		\citet{yagiSeto2017}. Millihertz detectors include LISA, TianQin and Taiji and their sensitivity curves were calculated using \citet{robsonCornishLiu2019} with inputs from \citet{luoetal2016}, \citet{huWu2017} and \citet{gongLuoWang2021}. 
	}
\end{figure}

In this subsection, we assess the detectability of the emitted GWs. One common property of magnetars is that they are observed to rotate slower than most of the NS population, with spin frequencies $\nu \lesssim 1~\text{Hz}$. On Earth, this is an issue as seismic noise limits our sensitivity to GW frequencies less than $\sim 10~\text{Hz}$. This means if the trapped ejecta model is to be used for magnetars or slowly rotating pulsars \citep[e.g.][]{tanetal2018, calebetal2022}, then any detectable signal must come from future space-based detectors. Of most relevance would be decihertz detectors like DECIGO \citep{setoKawamuraNakamura2001, kawamuraetal2011, kawamuraetal2021} or the Big Bang Observer (BBO) \citep{crowderCornish2005, harryetal2006, yagiSeto2011} but it is worth exploring whether millihertz space-based detectors like LISA \citep{amaro-seoaneetal2017, amaro-seoaneetal2023}, TianQin \citep{luoetal2016} or Taiji \citep{huWu2017} could also assist with detection. The sensitivity curves of the aforementioned space-based detectors are shown in Fig.~\ref{fig: space_based_sensitivity_curves}.

One of the most widely used measures of detectability is the SNR and the usual CW reference is \cite{jaranowskiKrolakSchutz1998}. There, the authors meticulously calculate the SNR of a quasi-infinite CW, emitted at once and twice the spin frequency of the source, with a fixed sky position and orientation. Due to the nature of GWs, there is a certain response function or ``antenna pattern'' associated with GW detection which captures the effect of GW detectors being less sensitive to certain areas in the sky for a particular orientation of the GW source. In this early study, we choose to omit this effect. As a result, any SNRs calculated here will be optimistic and the SNR in reality will likely be smaller. Alternatively, one could assume that the source has an unknown sky position and orientation, and so we would average over these variables. In this scenario, the SNRs presented here would be smaller by a factor of $\approx2/5$ \citep{jaranowskiKrolakSchutz1998}. However, since we will be primarily concerned with targets with known sky location, it makes little sense to do this (sky) averaging.
	
With all that said, we are still able to get a \textit{rough} idea of the detectability and so we follow \cite{jones2010} in defining the SNR as
\begin{align}
\label{SNR}
\rho \equiv \frac{h_0 \sqrt{T_\text{obs}}}{\sqrt{S_\text{n}(f)}}~,
\end{align}
where $T_\text{obs}$ is the total time spent observing the CW and $\sqrt{S_\text{n}(f)}$ is the amplitude spectral density of a given GW detector as a function of GW frequency $f$. $h_0$ is the GW amplitude and is defined as
\begin{align}
\label{def_h_0}
h_0 \equiv \langle h_{+}^2 + h_{\times}^2 \rangle^\frac{1}{2}~,
\end{align}
where the angled brackets represents a time average and $h_+$ and $h_\times$ are the amplitudes of the two polarisations. Equation~(\ref{def_h_0}) tells us that $h_0$ is the time-average of the root-sum-square of the two polarisations. The SNR is a statistic used to determine how much a signal stands out from the GW detector noise. Past a certain threshold, there is enough confidence to claim a detection, with the usual threshold being $\rho_\text{thres} = 11.4$ for conventional CWs corresponding to a 1 per cent false alarm rate and a 10 per cent false dismissal rate \citep{abbottetal2004threshold}.

In our case, the CW is not continuous but transient, lasting some finite duration \citep[e.g.][]{prixGiampanisMessenger2011}. This changes the SNR calculation depending on whether we assume the GW amplitude $h_0$ evolves during the emission period. If $h_0$ remains constant, then we simply replace $T_\text{obs}$ with $T_\text{GW}$, the duration for which GWs are emitted. If $h_0$ exponentially decays on time-scale $\tau_\text{GW}$, then we must change $T_\text{obs}$ into $\tau_\text{GW}/2$ \citep[e.g.][]{yimJones2020}. In practice, \citet{moraguesetal2023} found that we are allowed to make these substitutions between rectangular and exponential windows so long as the GW data are not overly interrupted during the emission period. It is also computationally much cheaper to conduct searches using rectangular windows and the maximum mismatch has been found to be acceptable so long as the duration of the rectangular window is not much longer than the signal duration \citep{prixGiampanisMessenger2011, keiteletal2019}. For the reasons above and in the absence of a detailed model of the trapping process, we will assume the GW amplitude to be constant during the entire emission period. The detection threshold for the SNR also changes for transient CWs \citep[e.g.][]{tenorioetal2022} but for simplicity, we use the conventional CW detection threshold of 11.4.


Following \citet{jones2010}, we split $h_0$ into two parts, one representing the $\dot{\phi}$ radiation and the other the $2\dot{\phi}$ radiation. Then, after substituting equations~(\ref{h_plus}) and (\ref{h_cross}) into equation~(\ref{def_h_0}), we find
\begin{align}
h_{0, 2\dot{\phi}} &= \frac{A}{\sqrt{2}} \sin^2\alpha \left[(1 + \cos^2\iota)^2 + 4\cos^2\iota\right]^{\frac{1}{2}}~, \\
h_{0, \dot{\phi}} &= \frac{A}{\sqrt{2}} \sin\alpha \cos\alpha \left[\sin\iota(1 + \cos^{2}\iota)^\frac{1}{2}\right]~,
\end{align}
where, for readability, we define
\begin{align}
A \equiv \frac{2}{d}\frac{G}{c^4} \dot{\phi}^2 |\Delta I_\text{d}|~.
\end{align}
If we substitute in $\Delta I_\text{d}$ for our model, we get
\begin{align}
\label{A_for_trapped_ejecta_model}
A = - \frac{2}{d}\frac{G}{c^4} \dot{\phi}^2 M_0 R_0^2 \left(\frac{\Delta \nu}{\nu_0}\right) \mathcal{G}~,
\end{align}
where, for known $\alpha$, we must have $l < l_\text{crit}$ for a glitch and $l > l_\text{crit}$ for an antiglitch. For known $l$, we must have $\alpha < \alpha_\text{crit}$ for a glitch and $\alpha > \alpha_\text{crit}$ for an antiglitch.

Putting the GW amplitudes into equation~(\ref{SNR}), we get
\begin{align}
\label{SNR_2phidot}
\rho_{2\dot{\phi}} &= \frac{A}{\sqrt{2}} \frac{\sqrt{T_\text{GW}}}{\sqrt{S_\text{n}(2\nu)}} \sin^2\alpha \left[(1 + \cos^2\iota)^2 + 4\cos^2\iota\right]^{\frac{1}{2}}~, \\
\label{SNR_phidot}
\rho_{\dot{\phi}} &= \frac{A}{\sqrt{2}} \frac{\sqrt{T_\text{GW}}}{\sqrt{S_\text{n}(\nu)}} \sin\alpha \cos\alpha \left[\sin\iota(1 + \cos^{2}\iota)^\frac{1}{2}\right]~.
\end{align}
In general, the SNR is larger when $A$ is larger, corresponding to sources that are nearer, rotating faster (but only to a certain point, as the GW detectors start becoming less sensitive) or sources with larger glitches/antiglitches. Also, the SNR increases with the square root of $T_\text{GW}$ as expected, but the value of $T_\text{GW}$ may be identified from observations which would prevent it from being a free parameter. 

Unless $\alpha$ is sufficiently small, at large $l$, only antiglitches will be observed and the EOS no longer has much influence. Interestingly, at large $l$, the SNR also becomes independent of $l$, since the factors in the numerator and denominator of $\mathcal{G}$ (equation~(\ref{def_G})) cancel. This can be seen below where we have taken the $l\rightarrow \infty$ limit for $A$
\begin{align}
A = - \frac{4}{5d} \frac{G}{c^4} \dot{\phi}^2 M_0 R_0^2 \left(\frac{\Delta \nu}{\nu_0}\right) \frac{1}{\sin^2\alpha}~,
\end{align}
which, when combined into the expression for the SNR for the $2\dot{\phi}$ radiation, we find 
\begin{align}
\rho_{2\dot{\phi}} = - \frac{2\sqrt{2}}{5d} \frac{G}{c^4} \dot{\phi}^2 M_0 R_0^2 \left(\frac{\Delta \nu}{\nu_0}\right) \frac{\sqrt{T_\text{GW}}}{\sqrt{S_\text{n}(2\nu)}} \left[(1 + \cos^2\iota)^2 + 4\cos^2\iota\right]^{\frac{1}{2}}~,
\end{align}
which is independent of $\alpha$, $l$ and $\gamma$. Similarly, the $l \rightarrow \infty$ limit for the $\dot{\phi}$ radiation gives
\begin{align}
\rho_{\dot{\phi}} = - \frac{2\sqrt{2}}{5d} \frac{G}{c^4} \dot{\phi}^2 M_0 R_0^2 \left(\frac{\Delta \nu}{\nu_0}\right) \frac{\sqrt{T_\text{GW}}}{\sqrt{S_\text{n}(2\nu)}} \cot\alpha \left[\sin\iota(1 + \cos^{2}\iota)^\frac{1}{2}\right]~,
\end{align}
which is independent of $l$ and $\gamma$, but not $\alpha$. 

In the $l \rightarrow 0$ limit, $A$ becomes
\begin{align}
A = - \frac{2}{d}\frac{G}{c^4} \dot{\phi}^2 M_0 R_0^2 \left(\frac{\Delta \nu}{\nu_0}\right) \frac{1}{\frac{5}{2}\sin^2\alpha + (2\gamma - 1)}~,
\end{align}
and the SNRs for both the $2\dot{\phi}$ and $\dot{\phi}$ radiation remain dependent on $\alpha$ and $\gamma$. This recovers the property that the EOS becomes more important for smaller $l$. Recall that if $\alpha < \alpha_\text{crit}$, then this limit on $l$ always gives a glitch.

\subsection{An example application}

At this point, we will fix some model parameters and then evaluate the results for the SNR at these two extreme values of $l$. Firstly, we choose the most optimistic viewing scenario for each of the two radiation frequencies by setting $\iota = 0\text{\textdegree}$ for the $2\dot{\phi}$ radiation and $\iota = 90\text{\textdegree}$ for the $\dot{\phi}$ radiation. This makes the square brackets in equations~(\ref{SNR_2phidot}) and (\ref{SNR_phidot}) equal $8$ and $1$, respectively.

We also choose to fix $\alpha = 30\text{\textdegree}$ and $\gamma = 0$ which corresponds to an $n=1$ polytrope. One can verify that both glitches and antiglitches are possible for this combination of parameters by selecting an appropriate value of $l$, as can be seen in Fig.~\ref{fig: ns_glitch_antiglitch_parameter_space}. We also use the canonical values of $M = 1.4~\text{M}_\odot$ and $R = 10~\text{km}$.

The remaining inputs are given by observations and, as an example, we use values for the most recent SGR J1935+2154 antiglitch and glitch. The relevant inputs for the antiglitch come from \citet{younesetal2023} and are $\Delta \nu / \nu_0 = - 5.8 \times 10^{-6}$, $T_\text{GW} = 4~\text{d}$ and $\dot{\phi} \approx \Omega = 1.94~\text{rad~s}^{-1}$. For the glitch, we assume the same $T_\text{GW}$ and $\dot{\phi}$ but take $\Delta \nu / \nu_0 = 6.4 \times 10^{-5}$ from \citet{geetal2022}. The true distance to the source is still under debate \citep[e.g.][]{kothesetal2018, zhouetal2020, bailesetal2021} but we will follow \citet{younesetal2023} by using a conservative estimate of $d = 9~\text{kpc}$ given by \citet{zhongetal2020}.

As for the GW detector sensitivity, we use the DECIGO sensitivity curve which, from Fig.~\ref{fig: space_based_sensitivity_curves}, tells us that $\sqrt{S_\text{n}(2\nu)} \approx \sqrt{S_\text{n}(\nu)} \approx 3\times 10^{-24}~\text{Hz}^{-\frac{1}{2}}$.

For the antiglitch, we take $l \rightarrow \infty$ which results in SNRs of $\rho_{2\dot{\phi}} = 5.7\times 10^{-5}$ and $\rho_{\dot{\phi}} = 3.5 \times 10^{-5}$. For the glitch, we take $l \rightarrow 0$ which results in SNRs of $\rho_{2\dot{\phi}} = 1.0\times 10^{-3}$ and $\rho_{\dot{\phi}} = 6.4 \times 10^{-4}$. Clearly, these values are far below the detection threshold so for the cases of ejecta held just off the surface, or at a height far above the surface, there are unlikely going to be any GWs detected.

However, and most crucially, this is not the end of the story. One can see that in the denominator of $\mathcal{G}$, there is an asymptote whenever $\alpha = \alpha_\text{crit}$ or $l = l_\text{crit}$. This leads to the main results of this model which is that, because of this asymptote, it is \textit{possible to get a detectable SNR ($\rho > \rho_\mathrm{thres}$) so long as $\alpha$ or $l$ is sufficiently close to its critical value}. Of course, we cannot interpret glitches or antiglitches if we sit exactly on the critical values for $\alpha$ or $l$, as then $\Delta I / I_0 = 0$ by equation~(\ref{change_in_MoI_neutron_star}), but the fact that we would have observed a glitch or antiglitch provides enough evidence for why $\alpha \ne \alpha_\text{crit}$ and $l \ne l_\text{crit}$.

However, this is not without its conditions. As we approach the critical boundary curve, the value of the square brackets in equation~(\ref{change_in_MoI_neutron_star}) becomes smaller which then forces $\delta M / M_0$ to be correspondingly larger in order to keep $\Delta I/I_0$ and hence $\Delta \nu/\nu_0$ constant, which is fixed by observations. Therefore, to ensure consistency, one needs to check whether $\delta M / M_0 < 1$. If this condition is not satisfied, then it means that it is not possible to detect GWs emitted from the source, even in the most optimal configuration for explaining GWs and glitches/antiglitches. This finding also explains why the SNR improves as one approaches the critical boundary curve, as more mass is required to be trapped which naturally gives off a stronger GW signal.

One can show this is exactly the case for SGR J1935+2154. For the above model parameters, we find $l_\text{crit} / R_0 \approx 0.2649$ using equation~(\ref{l_crit_neutron_star}). One finds that for detectable GWs ($\rho > 11.4$), $l$ needs to be within 4 decimal places of $l_\text{crit}$ making this an extremely finely tuned solution. This raises a conceptual issue if GWs are ever observed to be coincident with a glitch or antiglitch from SGR J1935+2154 as one might ask why there is a preferred ejecta height of $\approx l_\text{crit}$. In essence, it means that GWs will be detected along each curve of Fig.~\ref{fig: ns_glitch_antiglitch_parameter_space}. An incorrect conclusion to draw would be that a coincident GW detection from SGR J1935+2154 would allow us to constrain $\gamma$ to a high precision if this model is to be believed. When one checks whether $\delta M / M_0 < 1$ is satisfied using equation~(\ref{change_in_MoI_neutron_star}), one finds that it is not. The correct conclusion would be that the finely tuned solution is due the glitch/antiglitch data not being consistent with the supposedly detectable GWs. In other words, with the model inputs above, it is not possible to detect GWs coincident with glitches or antiglitches from SGR J1935+2154. This is not to say that it is impossible for all magnetar glitches/antiglitches, it just means that there needs to be the correct conditions, e.g.~a nearby, fast rotating magnetar with large glitches or antiglitches, for the correct interpretation of the results using the trapped ejecta model.

\section{Summary and Discussion}
\label{section_summary_and_discussion}

In this paper, we have put forward and examined a new model for transient CWs from magnetar glitches and antiglitches. At the same time, the toy model is able to naturally explain glitches and antiglitches using simple angular momentum arguments. The model proposes that during a glitch or antiglitch event, mass is ejected from the magnetar which is subsequently trapped at a height $l$ above the magnetar surface, at an inclination angle $\alpha$ from the rotation axis. The decrease in the magnetar's moment of inertia due to the mass loss has the effect of spinning up the system whereas the presence of the trapped ejecta has the opposite effect. These two effects are in contest with each other and whichever dominates dictates whether a glitch or antiglitch occurs. For a given EOS and ejected mass fraction, we are able to plot the glitch size as a function of $l$ and $\alpha$, which reveals a critical boundary curve separating glitches and antiglitches.

The presence of the trapped ejecta causes the magnetar system to undergo free precession and creates a time-varying mass quadrupole moment leading to GW radiation. We evaluated the GW radiation emitted and provided calculations for the GW luminosity, torque, energy and the SNR for GW detection. We found that the SNR improves as one approaches the critical boundary curve in $(l, \alpha)$ parameter space, due to more mass being trapped, subject to the condition $\delta M / M_0 < 1$. The SNR also improves for magnetars that are: closer, rotating faster, have larger glitches/antiglitches or if they trap the ejecta for longer. 

There appears to be a slight conflict when using the trapped ejecta model to interpret GWs coincident with glitches or antiglitches as it is harder to produce GWs for values of $(l, \alpha)$ that are more suited to explain glitches/antiglitches. However, if we are ever able to detect GWs coincident with glitches/antiglitches, then one can make some interesting conclusions if we accept that the model presented here is correct. A successful GW detection would imply the model parameters must fall into a very narrow range offering a tight constraint on the EOS (from parameter $\gamma$). However, this would be in contention with the glitch/antiglitch data, as one would need an abnormally large mass to be ejected. This could be resolved by invoking an additional glitch or antiglitch mechanism, or if we still accept the trapped ejecta as the sole reason for the glitch/antiglitch, then one would have to ask why and how such an abnormally large amount of mass is ejected. Ultimately, the model is still only a toy model and hence it may be premature to interpret any conflicts between GW and glitch/antiglitch data. It is instructive that more realistic physics is added to the model to help resolve potential future tensions.

One consequence of the model is that we would expect precession after a magnetar glitch or antiglitch, which could be observed electromagnetically as modulations in the pulse profile, polarisation and/or timing, given the precession lasts long enough. More work would need to be done on this in the context of this model, but this problem has already been tackled in quite a generic way by \citet{jonesAndersson2001} and \citet{gaoetal2023}.

There are, of course, many improvements that could be made to this model. There were several simplifying assumptions especially in the treatment of the trapped ejecta and the EOS. Future works should focus on improving this. Once this is done, it will be possible to tackle the inverse problem to reveal what can be learnt about the EOS in the event of a detection or non-detection of a coincident GW. To make the current toy model fully consistent with electromagnetic observations from \citet{younesetal2023} and \cite{geetal2022}, it will be important to include a mechanism to generate post-glitch FRBs. This too should be included in the future.

\section*{Acknowledgements}

The authors would like to thank the anonymous referee whose comments undoubtedly improved the clarity of this work. This work was supported by the National Natural Science Foundation of China (12247180, 11975027, 11991053), the National SKA Program of China (2020SKA0120300, 2020SKA0120100), the National Key R\&D Program of China (2017YFA0402602), the Max Planck Partner Group Program funded by the Max Planck Society, and the High-Performance Computing Platform of Peking University.

\section*{Data Availability}

The data used in this article was cited accordingly in the main text.



\bibliographystyle{mnras}
\bibliography{references} 




\appendix

\section{Mass-radius relation for polytropes}
\label{appendix_A}

In this appendix, we will derive the mass-radius relation for self-gravitating stars which are described by a polytropic equation of state (see also Section~3.3 of \cite{shapiroTeukolsky1983} or Section~13.3.2 of \cite{thorneBlandford2017}). By definition, polytropes are non-dissipative and hence barotropic, where pressure $P$ is a function of mass density $\rho$ only, and the pressure is a power law of the mass density
\begin{align}
P \equiv \kappa \rho^{\Gamma} \equiv \kappa \rho^{1 + \frac{1}{n}}~,
\end{align}
where $\kappa$ is a constant of proportionality and the second equality defines the polytropic index $n$ in terms of the adiabatic index $\Gamma = 1 + 1/n$. 

Non-rotating NSs in equilibrium are spherically symmetric so the pressure, mass density and gravitational potential $\Phi$ are all functions of the radial position $r$ only. In Newtonian gravity, NSs in equilibrium are governed by the equation of hydrostatic equilibrium
\begin{align}
\frac{dP}{dr} = - \rho \frac{d\Phi}{dr}~,
\end{align}
and Poisson's equation
\begin{align}
\frac{1}{r^2} \frac{d}{dr}\left(r^2 \frac{d\Phi}{dr}\right) = 4 \pi G \rho~.
\end{align}
Combining the two, we get
\begin{align}
\label{ode_polytropes}
\frac{1}{r^2} \frac{d}{dr}\left(\frac{r^2}{\rho} \frac{dP}{dr}\right) = - 4 \pi G \rho~.
\end{align}
From here, one can express the non-linear second order differential equation in terms of the dimensionless variables
\begin{align}
\theta^n &\equiv \frac{\rho}{\rho_\text{c}}~,\\
\xi &\equiv \frac{r}{a}~,
\end{align}
where $\rho_\text{c}$ is the density in the centre of the NS and $a$ is defined as
\begin{align}
a \equiv \left[\frac{(n + 1)\kappa}{4\pi G}\right]^{\frac{1}{2}} \rho_\text{c}^{\frac{1-n}{2n}}~.
\end{align}
Changing the variables of equation~(\ref{ode_polytropes}) from $P \rightarrow \rho \rightarrow \theta$ and $r \rightarrow \xi$, one gets
\begin{align}
\frac{1}{\xi^2} \frac{d}{d\xi}\left(\xi^2 \frac{d\theta}{d\xi}\right) = - \theta^n~,
\end{align}
which is known as the Lane-Emden equation for stellar structure and is used for self-gravitating bodies described by polytropic index $n$. This equation can be solved for $\theta(\xi)$ analytically for $n = 0$, $1$ and $5$ but for the most part, it needs to be solved numerically subject to the boundary conditions
\begin{align}
\theta (0) = 1~, \\
\frac{d\theta}{d\xi}(0) = 0~,
\end{align}
where the first boundary condition is the statement that the density at the centre of the star is $\rho_\text{c}$ and the second boundary condition is the requirement that the pressure gradient at the centre of the star must be zero (otherwise there would be a pressure force acting on an infinitely small mass as $\xi \rightarrow 0$). 

When solving the differential equation numerically, there will be a value for $\xi = \xi_1$ where $\theta = 0$, i.e. $\theta(\xi_1) = 0$. This defines the surface of the NS since this is when the density and pressure equals zero. The radius is then simply
\begin{align}
\label{radius_polytrope}
R = a \xi_1 = \left[\frac{(n + 1)\kappa}{4\pi G}\right]^{\frac{1}{2}} \rho_\text{c}^{\frac{1-n}{2n}} \xi_1~.
\end{align}
The mass is defined in the usual way
\begin{align}
M = 4\pi \int_{0}^{R} \rho r^2 dr~,
\end{align}
but one can use the Lane-Emden equation to more easily solve the above integral. To do this, we first change variables
\begin{align}
M = 4\pi \rho_\text{c} a^3 \int_{0}^{\xi_1} \theta^n \xi^2 d\xi~,
\end{align}
and substituting in the Lane-Emden equation, we get
\begin{align}
M = - 4\pi \rho_\text{c} a^3 \int_{0}^{\xi_1} \frac{d}{d\xi}\left( \xi^2 \frac{d\theta}{d\xi}\right)d\xi~.
\end{align}
The integral is now in the form that can be solved by simply invoking the fundamental law of integration and in doing so, we find the mass to be
\begin{align}
\label{mass_polytrope}
M = - 4\pi \left[\frac{(n + 1)\kappa}{4\pi G}\right]^{\frac{3}{2}} \rho_\text{c}^{\frac{3-n}{2n}}  \xi_1^2 \left.\frac{d\theta}{d\xi}\right|_{\xi_1}~.
\end{align}
Finally, we can combine equations~(\ref{radius_polytrope}) and (\ref{mass_polytrope}) by eliminating $\rho_\text{c}$ which gives us the mass-radius relation for polytropes
\begin{align}
M = - 4\pi R^{\frac{3-n}{1-n}} \left[\frac{(n + 1)\kappa}{4\pi G}\right]^{-\frac{n}{1-n}} \xi_1^{-\frac{1+n}{1-n}} \left.\frac{d\theta}{d\xi}\right|_{\xi_1}~.
\end{align}
We see that $M \propto R^{\frac{3-n}{1-n}}$ which is what was also found in \cite{shapiroTeukolsky1983} and \cite{thorneBlandford2017}, however, we note that the exponents derived here on the square brackets and $\xi_1$ differ from \cite{shapiroTeukolsky1983} and \cite{thorneBlandford2017}, which themselves are different from one another. This is even with the expressions for $M$ and $R$ all agreeing between the three calculations. We believe the other two references made a small typographic error when writing their manuscripts and the mass-radius relation here is correct.


\bsp	
\label{lastpage}
\end{document}